\documentclass[9pt,twoside]{rmaa-rho}
\RMxAAtemplatetype{\RMxAA}

\def\aap{A\&A}

\def\apjs{ApJS}
             
\def\msun{M$_{\odot}$}
\def\rsun{R$_{\odot}$}
\def\porb{$P_{orb}$}
\def\kms{km~s$^{-1}$}

\usepackage{xspace}
\usepackage[utf8]{inputenc}
\usepackage{textgreek}
\usepackage{xcolor}
\usepackage{deluxetable}
\newcommand{\vcas}{V630\,Cas\xspace}

\newcommand{\Msun}{M$_{\odot}$\xspace}

\newcommand{\Msunyr}{M$_{\odot}$\,yr$^{-1}$\xspace}
\newcommand{\teff}{T_\mathrm{eff}}

\vol{61}
\pages{30-36}
\thisyear{2026}
\doi{\href{https://www.astroscu.unam.mx/rmaa/RMxAA..XX-X}{https://www.astroscu.unam.mx/rmaa/RMxAA..XX-X}}

\title{A Comprehensive Study of the long-period Cataclysmic Variable V630 Cassiopeiae}

\author[1]{I. Mora Zamora \orcidlink{0000-0001-8600-4798}}
\author[1]{G. Tovmassian \orcidlink{0000-0002-2953-7528}}
\author[1]{J. Echevarr\'ia \orcidlink{0000-0001-5960-3023}}
\author[2]{D. Belloni\orcidlink{0000-0003-1535-0866}}

\affil[1]{Universidad Nacional Aut\'onoma de M\'exico. Instituto de Astronom\'{i}a. A.P. 70-264, 04510. Ciudad de M\'exico, M\'exico.}
\affil[2]{International Centre of Supernovae (ICESUN), Yunnan Key Laboratory of Supernova Research, Yunnan Observatories, Chinese Academy of Sciences (CAS), Kunming 650216, China}

\leadauthor{Mora-Zamora et al.}
\smalltitle{LaTex Macro RMxAA}

\corres{Iv\'an Mora Zamora}
\email{imora@astro.unam.mx}

\received{May 15th, 2026}
\accepted{\today}

\license{Texto de la licencia aqu\'i}

\setbool{rho-abstract}{true} 
\setbool{rho-resumen}{true} 

\begin{abstract}
    We present a comprehensive review of V630 Cassiopeiae, which is a very long-period cataclysmic variable. Based on new observations combining spectroscopy and high-resolution photometry provided by \textsc{TESS} and precise distance data from \textsc{Gaia}, we confirm and extend previously described characteristics. The light curves show differential behaviors between different epochs, attributed to changes in the disk's brightness distribution, likely due to non-axial structures such as precessing eccentric disks. The spectroscopy reveals complex H$\alpha$ line profiles, with dual components and an emission distribution that suggests a large, asymmetric, and possibly precessing disk, rather than a classic shock structure in the matter-flow impact region. Doppler tomography indicates that the emission originates at the periphery of the disk, distributed in an asymmetric pattern rather than a compact point, complicating the interpretation of radial velocities and limiting their use for accurately determining masses. Through photometric and spectroscopic analysis, we established a donor mass estimate with an approximate value of 0.38 \msun~at the adopted distance of 2409 pc, with significant uncertainty related to distance and other systematic factors. Finally, using the parameters of \vcas, binary evolution simulations were conducted with \textsc{MESA} incorporating CARB magnetic braking prescription to identify an evolutionary sequence that explains its properties. 
\end{abstract}

\keywords{accretion, accretion disks --- binaries: spectroscopic --- 
stars: novae, cataclysmic variables --- stars: individual (V630 Cassiopeiae) --- 
techniques: radial velocities --- techniques: spectroscopic --- 
white dwarfs --- stars: evolution}

\begin{resumen}
    Presentamos una revisión exhaustiva de V630 Cassiopeiae, una variable cataclísmica de período muy largo. Basándonos en nuevas observaciones que combinan espectroscopia y fotometría de alta resolución proporcionadas por TESS y datos de distancia precisos de \textsc{Gaia}, confirmamos y ampliamos características descritas previamente. Las curvas de luz muestran comportamientos diferenciales entre distintas épocas, atribuidos a cambios en la distribución de brillo del disco, probablemente debido a estructuras no axiales como discos excéntricos en precesión. La espectroscopia revela perfiles complejos de la línea Hα, con componentes duales y una distribución de emisión que sugiere un disco grande, asimétrico y posiblemente en precesión, en lugar de una estructura de choque clásica en la región de impacto del flujo de materia. La tomografía Doppler indica que la emisión se origina en la periferia del disco, distribuida en un patrón asimétrico en lugar de un punto compacto, lo que complica la interpretación de las velocidades radiales y limita su uso para determinar con precisión las masas. Mediante análisis fotométricos y espectroscópicos, se estimó la masa del donante con un valor aproximado de 0.38 \msun~a la distancia adoptada de 2409 pc, con una incertidumbre significativa relacionada con la distancia y otros factores sistemáticos. Finalmente, utilizando los parámetros de \vcas, se realizaron simulaciones de evolución binaria con MESA, incorporando la prescripción de frenado magnético CARB para identificar una secuencia evolutiva que explique sus propiedades.
\end{resumen}

\begin{document}

\maketitle
\pagestyle{fancy}\thispagestyle{firststyle}


\section{INTRODUCTION}

\RMxAAstart{C}ataclysmic variables (CVs) are semi-detached binary systems in which a white dwarf (WD) accretes material from a main-sequence or slightly evolved donor star. The evolution of CVs from longer to shorter orbital periods above the period gap has traditionally been explained by magnetic braking, based largely on magnetically driven winds that carry away angular momentum \citep{Knigge_2011,Belloni+Schreiber_2023_review}. Although uncertainties remain about the form and strength of magnetic braking, growing evidence from recent observations and numerical modeling suggests that magnetic braking saturates in rapidly rotating low-mass main-sequence stars \citep{El-Badry_2022}.

New models incorporating magnetic saturation have been successfully applied to CVs \citep{Barraza-Jorquera_2025,Barraza-Jorquera_2026,Dodon_2026,Tang_2026}, their progenitors \citep{Belloni_2024a,Blomberg_2024,Thai_2026}, and related objects \citep{Whitebook_2026a,Whitebook_2026b}, which also indicate that in addition to saturation, disruption and boost are required to explain observations. Regarding evolved stars (i.e. subgiants and red giants), it has been shown that even more efficient magnetic braking is required to explain persistent low-mass X-ray binaries, persistent ultra-compact X-ray binaries, and the existence of short-period detached millisecond pulsar binaries \citep[e.g.][]{Van+Ivanova_2019,Van_2021,Soethe_2021,Deng_2021,Chen_2021,Shahbaz_2022,Wei_2023,CastroSegura_2024,Echeveste_2024,Cui_2024,Yang_2024,Kar_2025,Yang_2025}. Similar prescriptions applied to CVs with evolved donors can also solve the fine-tuning and hydrogen-abundance problems to form AM\,CVn binaries \citep{Belloni+Schreiber_2023}. In addition, this framework can explain the formation of short-period detached binaries hosting extremely low-mass WDs with more massive WD companions \citep{Aros-Bunster_2025,Belloni_2025}. This leads to a single evolutionary sequence from CVs with evolved donors to detached double WD binaries with orbital periods of a few hours \citep{El-Badry_2021} and AM\,CVn binaries. If this scenario is correct, the properties of CVs with evolved donors and longer orbital periods -- corresponding to an earlier stage of the same evolutionary sequence -- should be consistent with strong magnetic braking. Indeed, this scenario has been supported by observations of long-period CVs such as V479\,And and V1082\,Sgr \citep{Tovmassian_2026}, and SDSS\,J085210.48+783246.6 \citep{Tovmassian_2026}. The study of long-period CVs therefore provides a natural laboratory to test this scenario.

V630 Cassiopeiae is an atypical and poorly observed CV star with a remarkably long orbital period of about 2.5~days. Its distance \citep[Bailer-Jones catalog,][]{Bailer-Jones_2021,Bailer-Jones_2021Data} is about $2410^{+300}_{-194}$ pc corresponding to a median geometric distance (\texttt{rgeo}). V630 Cas was classified in the \citet{Duerbeck_1987} catalog as a likely WZ Sagittae-type dwarf nova, which is a specific subtype of CV stars characterized by relatively large-amplitude outbursts and relatively long intervals between these outbursts. Only three outbursts have been reported. The first observed outburst occurred in 1950 and lasted about 2 months \citep{Whitney_1973}, during which the star ranged from $m_{pg} = 17.1$ in quiescence to $m_{pg} = 12.3$ at maximum. A second outburst was observed in 1992 \citep{Honeycutt_1993}, featuring a 70-days increase to a peak brightness of $m_V = 14.3$, followed by a decline over approximately 30 days. The most recent outburst occurred in the first half of 2009 \citep{Shears_2010}. It lasted about 104 days, with the rise to maximum brightness taking 61 days, which was slightly slower than the 43-day decline. At its brightest, it reached a magnitude of $m_V = 14.0$, 2.3 magnitudes above the average quiescent magnitude. All three outbursts displayed faster declines than rises \citep{Shears_2010}. However, the 1950 outburst is distinguished from the other two by being brighter and having a faster ascent rate. Shears \& Poyner suggest that V630 Cas could be an as-yet-unrecognized member of the hybrid CN/DN group, although there is no evidence that a nova outburst has occurred. Other long-period CVs that exhibit similar patterns are BV Cen \citep[\porb = 0.611 days,][]{Bateson_1974}; GK Per \citep[\porb = 1.997 days,][]{Pezzuto_1996,Evans_2009} and V1129 Cen \citep[\porb= 0.893 days,][]{Bruch_2017}. This may also be a consequence of the long orbital period. 

The first spectroscopic study was conducted by \citet{Orosz_2001}. These authors found a long-period system with $P_{orb} = 2.56387$ days. Although they were able to observe both, the emission and absorption components, they find a semi-amplitude $K_2 = 132.9 \pm 4.0$ \kms~and a mass function $f(M_1) = 0.62 \pm 0.06$ \msun~from which they derive $M_1 = 0.98^{+0.17}_{-0.10}$ \msun~and $M_2 = 0.17^{+0.03}_{-0.01}$ \msun~based on an inclination angle of $i  = 74^{\circ}$, which was determined by fitting an ellipsoidal model to their light curve. They also found that the secondary star is a stripped giant. Although they quote a value for the semiamplitude of the emission lines of $K_1 = 39.1 \pm 4.9$ \kms~they do not rely on this value due to the errors given the large scatter seen in their radial velocity (RV) fit. As we show in Section~\ref{subsec:MassFuncWD}, however, this inclination -- and consequently the derived component masses -- require revision once an updated donor radius is taken into account; we adopt the revised values throughout the remainder of this paper.

\citet{Thorstensen_2017} report a new measurement of $P_{\text{orb}} = 2.56388(2)$ using the Hiltner 2.4m and McGraw-Hill 1.3m telescopes at MDM Observatory on Kitt Peak, Arizona, obtained from spectroscopy.

Here, we present new time-resolved optical spectroscopy of \vcas obtained at the Observatorio Astron\'omico Nacional at San Pedro M\'artir combined with \textsc{TESS} photometry. We refine the orbital period. Combining SED fitting of infrared photometry with the diagnostic diagram method, we revise the donor mass. Doppler tomography in H$\alpha$ and Fe\,{\sc i} confirms the detection of the donor star's absorption signal and reveals an azimuthally asymmetric accretion disc. We discuss the observed H$\alpha$ emission morphology. Finally, we consider the evolutionary track of the observed system. 

\begin{table}
\begin{center}
\caption{Log of observations of V630 Cas.}
\begin{tabular}{cccc} \hline
Spectroscopy & HJD & No. of &  Exposure \\
 night & 2450000+ & spectra & time (s) \\ \hline
2017-09-29 & 8025 & 24 &  900 \\ 
2017-09-30 & 8026 & 19 & 1200 \\ 
2017-10-01 & 8027 & 20 & 1200 \\ 
2017-10-11 & 8037 & 10 & 1200 \\ 
2017-10-12 & 8038 &  6 & 1200 \\ 
2017-10-13 & 8039 &  6 & 1200 \\ 
2017-11-26 & 8083 &  9 & 1200 \\ 
2017-11-27 & 8084 &  7 & 1200 \\
2018-01-20 & 8138 &  2 & 1200 \\ 
2018-01-21 & 8139 &  4 & 1200 \\ 
2018-01-22 & 8140 &  4 & 1200 \\
2018-01-23 & 8141 &  5 & 1200 \\
2018-01-24 & 8142 &  2 & 1200 \\
2018-01-25 & 8143 &  2 & 1200 \\
2018-01-26 & 8144 &  2 & 1200 \\
2019-01-19 & 8502 &  2 & 1200 \\
2019-01-20 & 8503 &  3 & 1200 \\
2019-12-10 & 8827 &  3 & 1200 \\
2019-12-12 & 8829 &  5 & 1200 \\
2019-12-13 & 8830 &  2 & 1200 \\
2019-12-14 & 8831 &  2 & 1200 \\
2019-12-15 & 8832 &  3 & 1200 \\ \hline
\end{tabular}
\label{tabla:Bitac}
\end{center}
\end{table}

\section{OBSERVATIONS}\label{sec:observations}

142 new spectra were obtained with the 2.1m telescope of the Observatorio Astron\'omico Nacional at San Pedro M\'artir, using the Boller \& Chivens spectrograph and a E2V42-40 2048 x 2048 CCD detector in the $5400 - 6700$ \r{A} range with $R\sim 1,200$ corresponding to a resolution of $\sim$~2.5~\r{A} FWHM. The log of observations is shown in Table \ref{tabla:Bitac}. The observations were conducted during the months of September, October, and November 2017, as well as in January 2018 and in January and December 2019, in a specific way to obtain as many orbital phases as possible. This is problematic as the orbital period is close to a multiple of 2.5 days and daily consecutive observations cover only restricted orbital phases. The exposure time for each spectrum was 1200 s, except for the first night for which we used 900 s. These spectra were reduced and treated with the Image Reduction and Analysis Facility software {\sc IRAF}\footnote{IRAF is distributed by the National Optical Astronomy Observatories, which are operated by the Association of Universities for Research in Astronomy, Inc., under cooperative agreement with the National Science Foundation.}. Using the same configuration and instruments, spectra of the standard star 61 Cyg A of spectral type K5 V \citep{Keenan_1989} were also obtained.

\begin{figure}
    \includegraphics[width=\columnwidth]{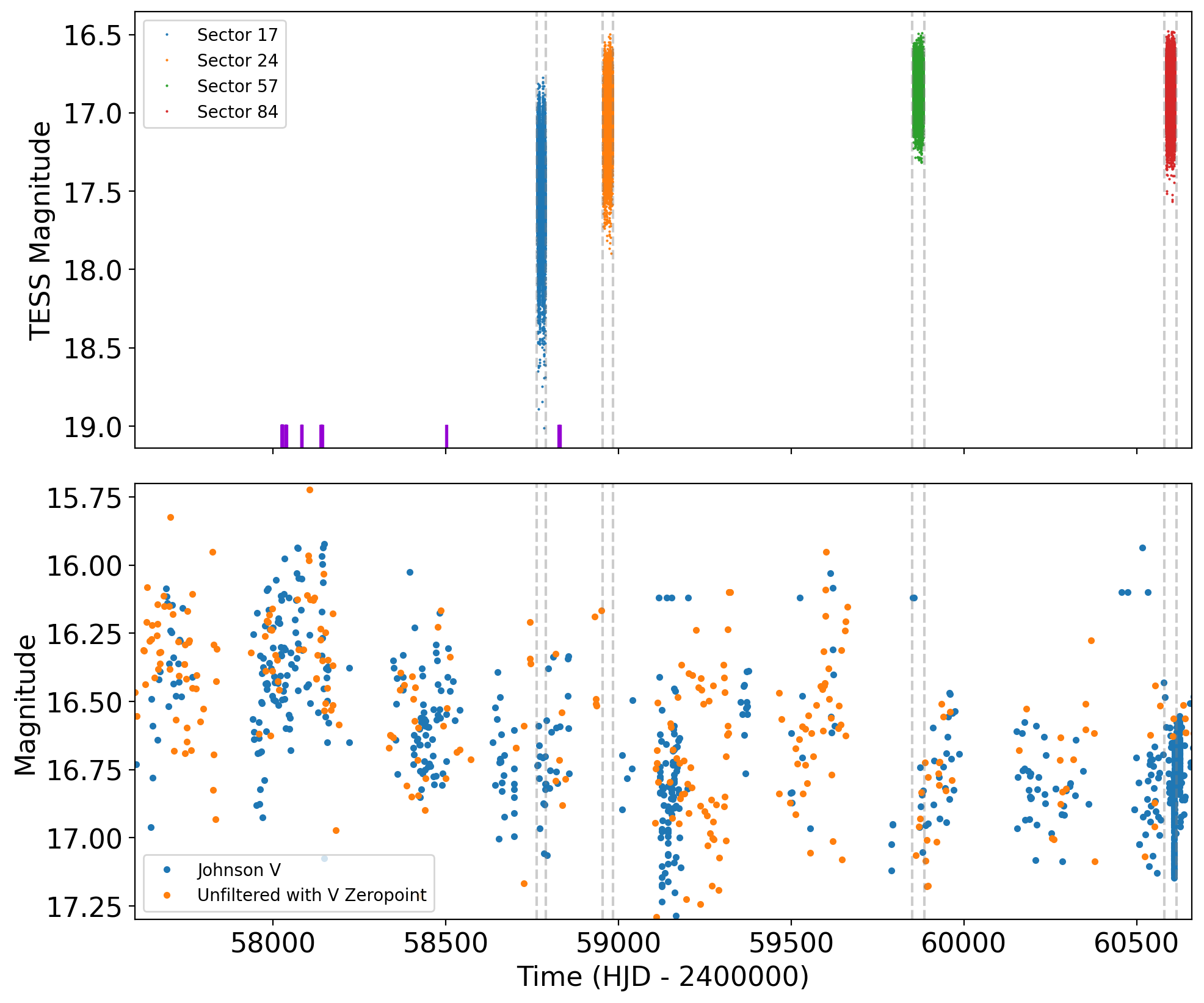}
    \caption{The global \textsc{TESS} light curve in the upper panel and epochs of spectroscopic observations marked at the bottom of the panel. In the bottom panel, the \textsc{AAVSO} CCD data demonstrate the brightness variability on the long term.}
    \label{fig:TESS_AAVSO}
\end{figure}

The spectra are of uneven quality. Of the 142 exposures, 111 have a signal-to-noise ratio S/N~$\geq 15$ per pixel measured in the continuum near H$\alpha$; the remaining 31 suffer from poor seeing, high airmass, or variable sky transparency and were excluded from the Doppler tomography analysis, though all 142 were retained for the radial-velocity fits. An additional source of systematic uncertainty is the wavelength calibration. The Boller \& Chivens spectrograph at SPM was modified to accommodate a heavy CCD camera, which introduced mechanical flexure: the zero-point of the wavelength scale drifts as the telescope tracks across the sky. This drift is corrected using the positions of bright night-sky emission lines, a procedure that is reliable when the target is observed continuously throughout a night, because the flex is gradual and well-sampled. However, for a system with \porb~$\approx 2.56$~d, each observing night covers only a fraction of the orbit, and the data were gathered across 22 separate runs spanning more than two years. The flex pattern differs from night to night, and systematic zero-point offsets of order tens of \kms\ between runs cannot be fully eliminated by sky-line corrections alone. As a result, our radial-velocity measurements carry a systematic uncertainty that dominates over the formal fitting errors, particularly at the quadrature phases most sensitive to $K_2$. For this reason, we adopt the $K_2$ value of \citet{Orosz_2001} as the primary constraint on the orbital dynamics throughout this paper, supplementing it with our own measurements where they provide independent confirmation.

\section{TESS DATA ACQUISITION AND PROCESSING}

\begin{figure}
    \includegraphics[width=\columnwidth]{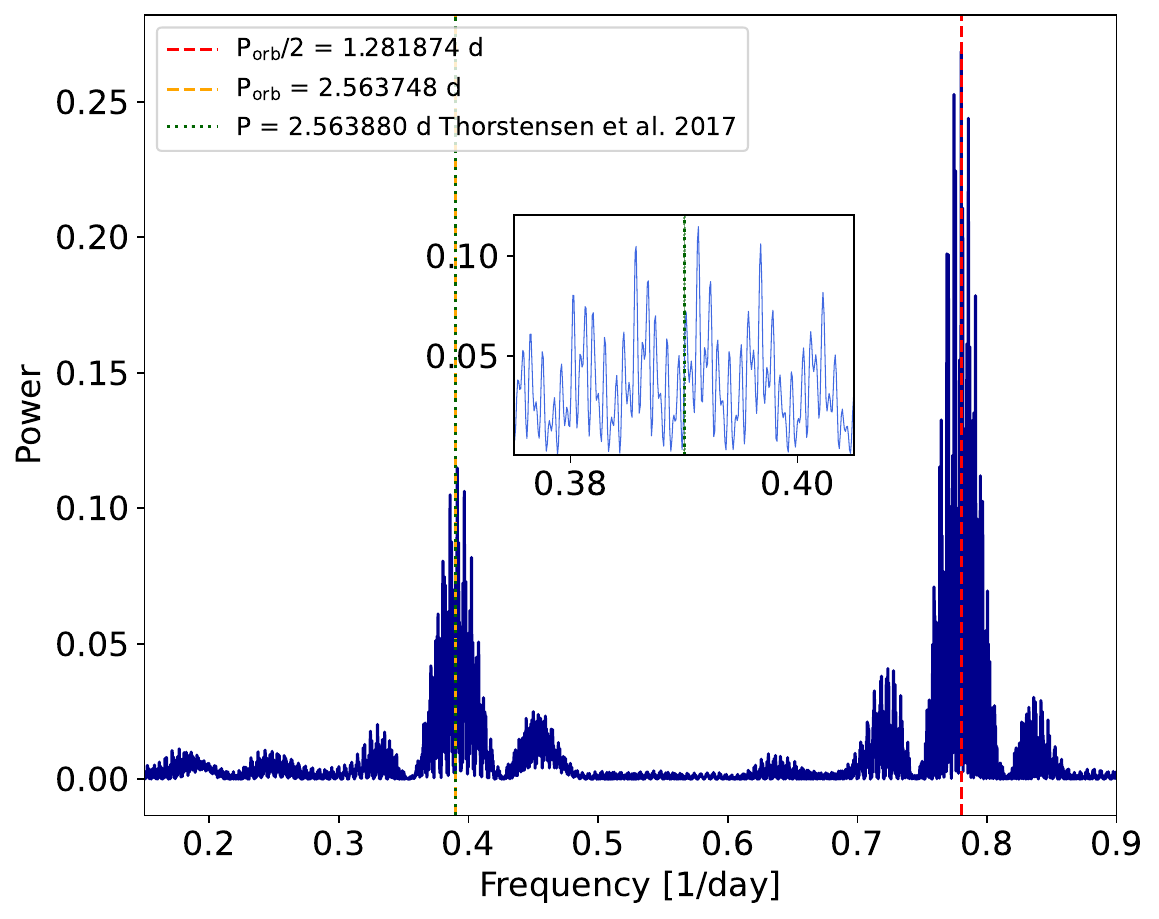}
    \caption{The Lomb-Scargle power plot for the \textsc{TESS} collective data.}
    \label{fig:TESS_lc}
\end{figure}

We retrieved photometric time-series data of V630~Cas from the Mikulski Archive for Space Telescopes (MAST), which provides calibrated light curves from the \textit{Transiting Exoplanet Survey Satellite} (\textsc{TESS}, \citet{Ricker_2015}). Specifically, we downloaded Simple Aperture Photometry (SAP) and Pre-search Data Conditioning SAP (PDCSAP) light curves from Sectors 17, 24, 57, and 84, spanning different observing epochs from 2019 to 2024. These data were accessed in FITS format using the \textsc{TESS} Input Catalog ID 431521404.

\textsc{TESS} light curves from MAST are preprocessed by the Science Processing Operations Center (SPOC), which applies background subtraction, cosmic-ray removal, and systematics-correction pipelines \citep{Jenkins_2016}. While PDCSAP flux generally offers stable light curves suitable for exoplanet detection, we found that both SAP and PDCSAP fluxes show long-term instrumental trends or discontinuities in some sectors. Such issues have been documented in the literature, particularly for bright or crowded-field targets \citep{Oelkers_2018}, and can appear as artificial slopes or offsets between light curve segments.

In Figure~\ref{fig:TESS_AAVSO}, we present a comparison between the \textsc{TESS} photometric light curves (Sectors 17, 24, 57, and 84) and long-term ground-based \textsc{AAVSO} monitoring of V630~Cas. For all sectors, we used the \texttt{PDCSAP\_FLUX} values from the FITS files, pre-conditioned to correct for instrumental systematics and crowding. The fluxes were converted to approximate magnitudes using the relation:

\[
m_{\mathrm{TESS}} = -2.5 \log_{10}(\texttt{PDCSAP\_FLUX}) + 22.5.
\]
A violet-colored marker at the bottom of the upper panel indicates the time of our spectroscopic observations (see Table\,\ref{tabla:Bitac}). 

\begin{figure*}
\includegraphics[width=\textwidth]{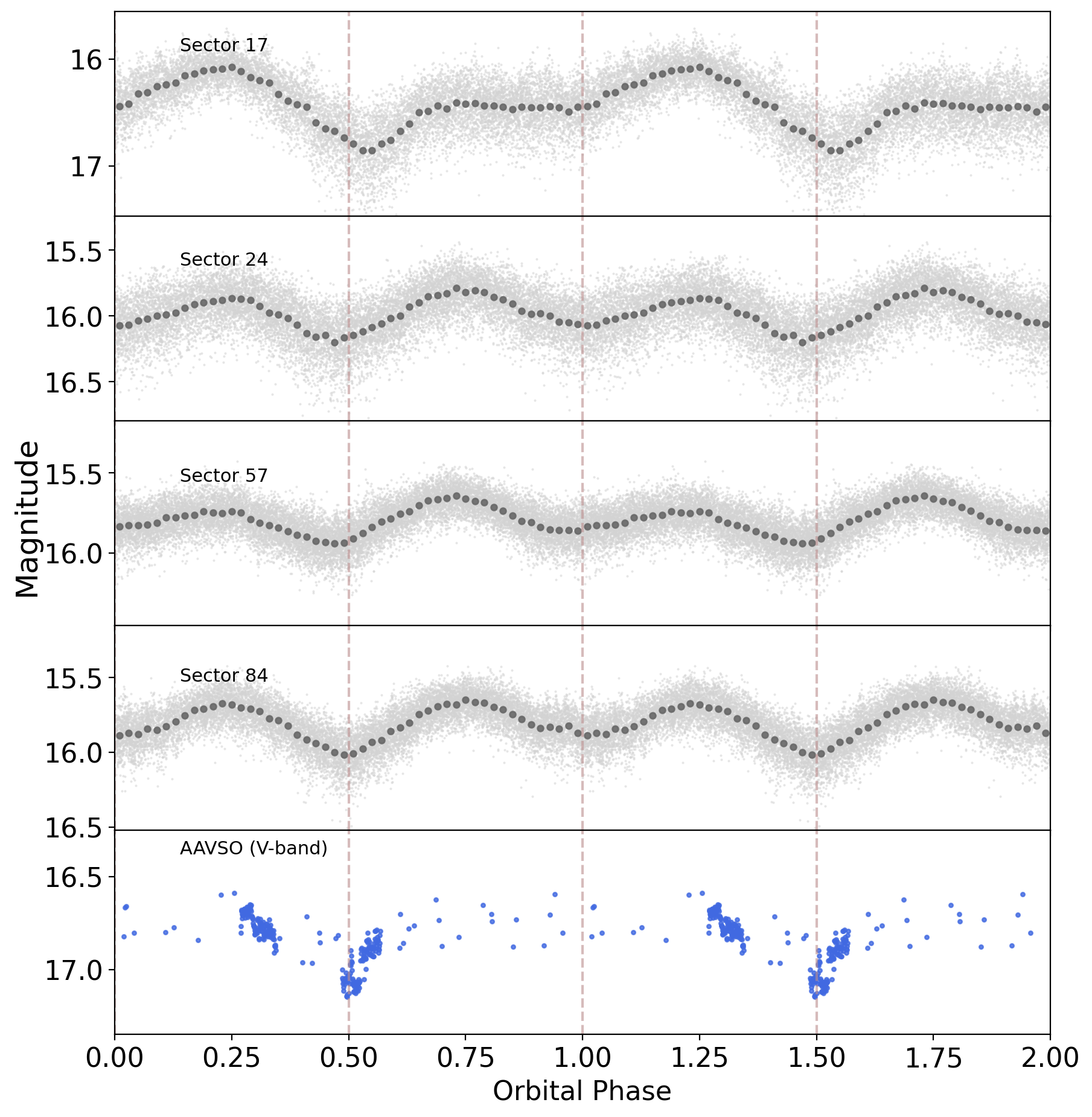}
    \caption{Phase--folded light curves of V630~Cas from \textsc{TESS} (Sectors~17, 24, 57, and~84) and contemporaneous \textsc{AAVSO} Johnson~$V$ observations, folded with the orbital period $P_{\mathrm{orb}} = 2.563886$~d and spectroscopic zero point $T_{0} = 2450709.76$. Each of the upper four panels shows one \textsc{TESS} sector: grey points indicate individual \texttt{PDCSAP\_FLUX} measurements converted to magnitudes, while dark-grey circles represent phase--binned averages ($\Delta\phi = 0.02$). Vertical dashed lines indicate orbital phases 0.0 and 0.5.  The bottom panel displays the \textsc{AAVSO} Johnson $V$ light curves filtered to the same time intervals as the \textsc{TESS} sectors, illustrating the consistency of long-term optical variability. Although the overall modulation is persistent, the amplitude and morphology of the double--humped pattern vary between sectors, suggesting that the variability arises primarily from changes in the accretion disk structure rather than from pure geometrical ellipsoidal effects.}
    \label{fig:tess_aavso}
\end{figure*}

The bottom panel of Figure \,\ref{fig:TESS_AAVSO} shows a collection of CCD-grade data in Johnson-V and unfiltered (shifted to the V zero point) bands to demonstrate a longer time-based behavior of the object. The system remained steadily brighter before falling by $\approx 0.25$ magnitudes around JD\,2458500, an epoch around which the observations reported here were performed.

To mitigate these calibration inconsistencies and enable robust temporal and period analyses, we applied a flux normalization strategy to each sector independently. We adopted a simple median-based normalization for sectors with relatively uniform flux (Sectors 17, 57, and 84): the median flux was divided out and unity subtracted to express each light curve as a fractional deviation. Sector~24 required additional treatment: the light curve was split at its temporal midpoint, the first segment was median-normalized, and the second segment---which showed a systematic instrumental trend---was divided by a second-order polynomial fit before the same subtraction was applied. This brought both segments to a common baseline.

\section{PERIOD ANALYSIS}\label{sec:PeriodAnalisis}

\subsection{Photometry}\label{subsec:photometry}

We performed a periodicity analysis using the Lomb–Scargle periodogram \citep{Lomb_1976, Scargle_1982}, which is well suited for unevenly sampled astronomical time series (see Figure \,\ref{fig:TESS_lc}). A dense frequency grid from 0.1 to 1.0\, cycles\,d$^{-1}$ was used to search for orbital and double-hump modulations commonly seen in interacting binaries.  

For the combined, normalized \textsc{TESS} dataset from Sectors~17, 24, 57, and~84, the periodogram reveals a dominant peak at P about 1.282 \,d (frequency $\approx 0.780$\,cycles\,d$^{-1}$). Because a tidally distorted, Roche-lobe-filling donor produces \emph{two} photometric maxima per orbit, the dominant photometric period is half the orbital period. Doubling yields $P_{\rm orb} = 2 \times 1.282 = 2.564$\,d, in excellent agreement with $P_{orb} = 2.56388(2)$\,d determined by \citet{Thorstensen_2017} and with our refined spectroscopic period (Section\,\ref{subsec:optim}). The two-humped signature is characteristic of ellipsoidal variability, and the \textsc{TESS} data covering multiple sectors confirm the long-term stability of this modulation.

Figure~\ref{fig:tess_aavso} shows the phase-folded light curves with the refined \porb~mentioned in Section \ref{subsec:optim}. The double-humped profile is present throughout the observations. However, its amplitude and shape vary between epochs—likely due to changes in the disk brightness distribution or mass-transfer rate, rather than a change in orbital geometry.  

The archival \textsc{AAVSO} photometry, primarily in the Johnson~$V$ band, shows consistent modulation at the same period, with no significant long-term deviations during the epochs of \textsc{TESS} monitoring. This agreement confirms that the variability is intrinsic to the system rather than an artifact of calibration differences between space- and ground-based datasets.

\begin{figure}
    \includegraphics[width=\columnwidth]{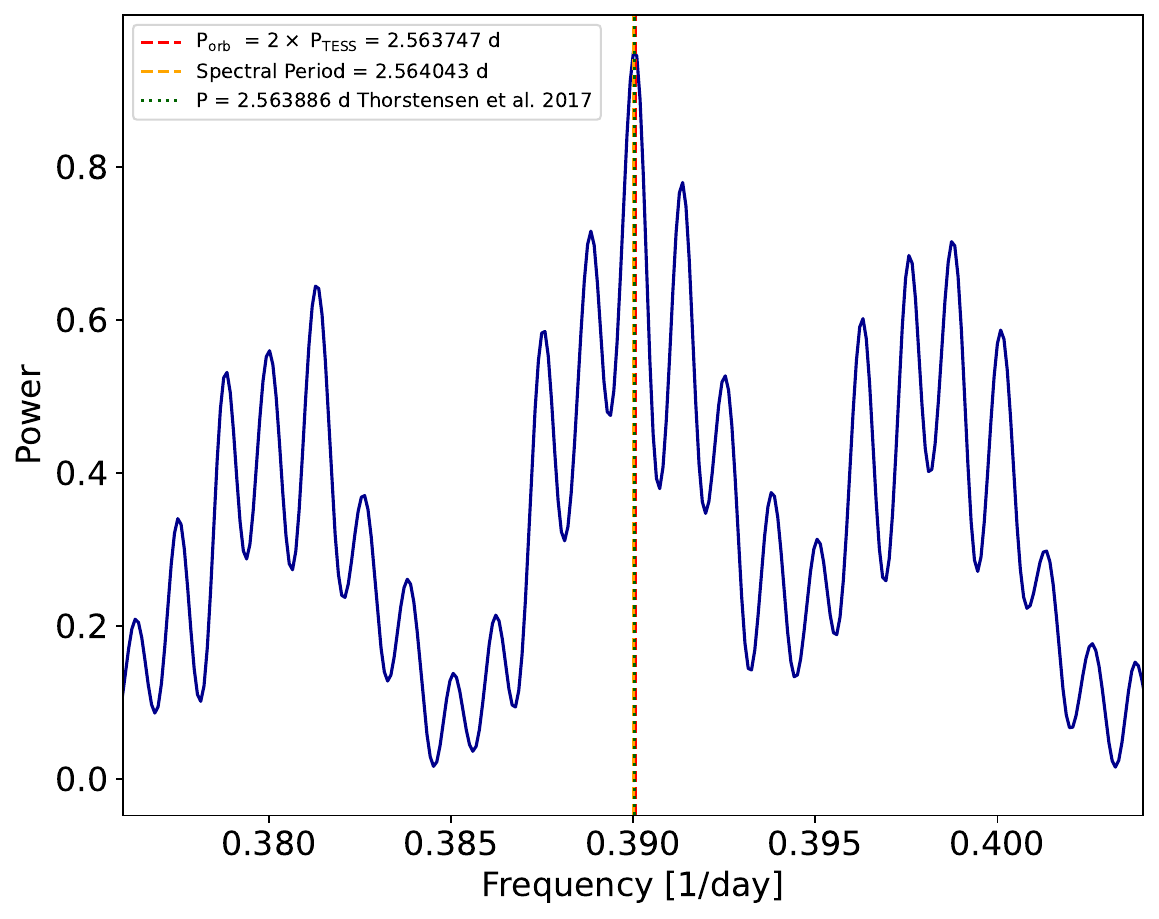}
    \caption{Power spectrum of the set of absorption features in the V630~Cassiopeiae spectra (see text).} 
    \label{fig:TESS_PS}
\end{figure}

The phase-folded light curves (Figure~\ref{fig:tess_aavso}) show a clear asymmetry between the two ellipsoidal minima. The \textit{deeper} minimum falls near phase~0.5 (superior conjunction, donor behind the disc and WD), while a \textit{shallower} feature appears near phase~0.0 (inferior conjunction, donor between us and the WD), as also noted by \citet{Orosz_2001}. This behavior is consistent with the accretion disc shielding the inner face of the donor at superior conjunction. Crucially, no deep eclipse is detected at phase~0.0 in any sector, meaning that the line of sight to the disc emission region is never fully intercepted — ruling out very high inclinations. \citet{Orosz_2001} derived an inclination in the range $66.96^{\circ} < i < 78.08^{\circ}$ by fitting an ellipsoidal model to their ground-based light curve. As the donor radius derived in Section~\ref{subsec:donormass} turns out to be significantly larger than that implicit in the Orosz et al. solution, this range cannot simply be carried forward unchanged; we revisit the determination of $i$, coupling it to the resulting white-dwarf mass, in Section~\ref{subsec:MassFuncWD}, once the donor parameters needed for that calculation have been established.

A further subtlety is that the minimum near phase~0.5 does not fall at exactly $\phi = 0.5$ in all sectors. In Sector~17, the deeper minimum is displaced by $\delta\phi \simeq 0.08$, while in Sectors~24, 57, and~84 it lies closer to the expected value. Because the orbital period is determined to six decimal places and all sectors share the same spectroscopic $T_0$, this cannot be a phasing artefact. Instead, it reflects a genuine epoch-to-epoch change in the azimuthal brightness distribution of the disc — consistent with a precessing or eccentric disc whose asymmetric emissivity pattern evolves on a timescale of years. This floating minimum provides independent photometric evidence for the disc asymmetry that also manifests in the H$\alpha$ profile morphology and the Doppler tomogram (Sections~\ref{sec:Halpha} and~\ref{sec:tomo}); we return to the anomalous depth of the Sector~17 minimum, quantitatively, in Section~\ref{subsec:MassFuncWD}.

\subsection{Spectroscopy}\label{subsec:Periodo-espect}

To obtain the radial velocities of the secondary star, we used a cross-correlation method using the metallic absorption lines of the donor (Fe\,{\sc i}, Ca\,{\sc i}, and similar features) in the spectral range 4900–5800 \AA, avoiding the emission lines. This method was applied with the {\sc fxcor} task in the {\sc iraf rv} package. For this method, a template spectrum with a reliable heliocentric RV is required. In this case, the standard star 61 Cyg A, mentioned in Section \ref{sec:observations}, which has an RV of $ 65.82 \pm 0.06$  \kms\,, was used \citep{Halbwachs_2018}. This method was applied to 140 observed spectra, excluding only two with extremely low signal-to-noise ratios. 

\begin{figure}
    \centering
    \includegraphics[width=\columnwidth]{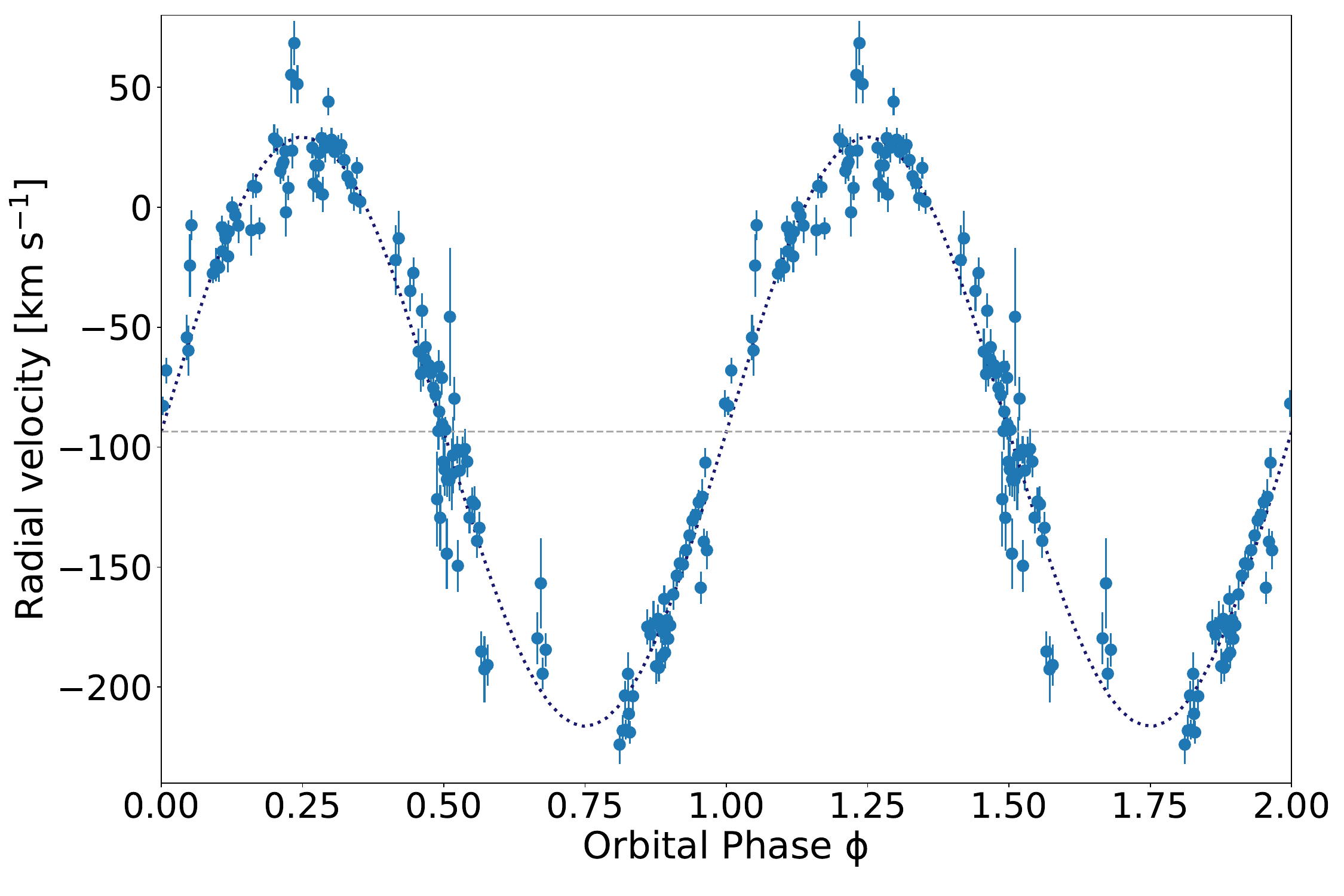}
    \caption{Radial-velocity curve of the secondary star in quiescence. The derived $K_2$ is $122.8 \pm 2.3$  \kms and $\gamma = -93.5$ \kms. The time of inferior conjunction and the orbital period used here are explained in the text.}
    \label{fig:vel_sec}
\end{figure}

Once the radial velocities were determined, we measured the orbital period based on our data by using the Lomb-Scargle power spectrum algorithm as implemented in Astropy \citep{Astropy_2022}. The resulting power spectrum is presented in Figure~\ref{fig:TESS_PS}. The highest peak in the power corresponds to 2.564043\, d. A nonlinear least-squares sinusoid fit to the same radial velocities, with $T_0$ fixed to the independently determined photometric epoch, gives $P = 2.5638 \pm 0.0012$\,d, statistically consistent at the $0.1\sigma$ level with the period of $2.56388(2)$\,d determined by \citet{Thorstensen_2017}. Subsequently, the RV curve was modeled using {\sc orbital}\footnote{Available at \url{https://github.com/Alymantara/orbital_fit}.}, a simple least-squares tool allowing to determine the orbital parameters, employing reduced $\chi^2$ as our goodness-of-fit parameter, to evaluate the orbital parameters for a circular orbit of the form:

\begin{equation}
V(t) = \gamma + K \sin\left(2\pi\frac{T - T_0}{P_{orb}}\right),
\label{radvel}
\end{equation}
where $\gamma$ is the systemic velocity, $K$ the semi-amplitude, $T_0$ the time of inferior conjunction of the donor, and \porb~is the orbital period (the refined orbital period as determined in Sec.\,\ref{subsec:optim}). The results of the fit are presented in Figure \ref{fig:vel_sec}. Our fit yields a value for $K_2= 122.8 \pm 15.0$ \kms~and a $\gamma=-93.5 \pm 5.0$ \kms.

The sinusoidal fit to the absorption-line velocities yields a formal statistical uncertainty of $\pm \ 2.3$\,\kms, but the data suffer from a systematic effect that dominates the error budget.  Inspection of the residuals reveals that at quadrature phases ($\phi \approx 0.68$--$0.77$, where sensitivity to $K_2$ is greatest), the scatter within a single night reaches 100\,\kms, far exceeding the orbital motion of $<5$\,\kms\ over the same time baseline. This is most naturally explained by variable veiling of the donor's absorption features by the accretion disc continuum. When the disc continuum is elevated, it raises the local spectral continuum and reduces the equivalent widths of the donor absorption lines. The cross-correlation peak becomes shallower and noisier, and the centroid velocity is less well-determined, increasing the scatter in measured $K_2$ without a systematic directional bias. Epochs of strong disc emission therefore contribute large residuals of either signal, while epochs with a fainter disc yield clean, well-defined cross-correlation peaks that track the donor's true orbital motion. The effect is epoch-dependent: on HJD\,58038, three consecutive spectra obtained over 25\,min are mutually consistent at $v \approx -185$ to $-193$\,\kms\ at $\phi \approx 0.76$, implying $K_2 \approx 130$--$135$\,\kms.  On the same orbital phase observed on other nights, the measured velocities span $-46$ to $-149$\,\kms, a range attributable entirely to varying disc contamination. Iterative sigma-clipping and phase-bin scatter weighting were applied but did not converge to a higher $K_2$, because both contaminated (slow) and uncontaminated (fast) spectra at the extrema carry comparable statistical weight and cancel in the fit.  We therefore estimate a systematic uncertainty of $\pm 15$\,\kms\ on $K_2$, which dominates over the formal fitting error.

Additional support for a higher $K_2$ comes from the independent study of \citet{Orosz_2001}, who obtained $K_2 = 132.9 \pm 4.0$\,\kms\ from spectra acquired at a time when the system was presumably in a different disc state.  Several arguments favour the Orosz value over our lower formal result.  First, their spectra were obtained at higher resolution and with a dedicated wavelength calibration strategy, whereas our data suffered from wavelength calibration drifts that were identified in post-processing and represent an additional source of systematic velocity error, particularly affecting the zero-point of individual nightly solutions. Second, the subset of our own data obtained when the disc continuum was evidently low (night HJD\,58038, where three consecutive measurements agree to within 4\,\kms) is fully consistent with $K_2 \approx 130$--$135$\,\kms, in good agreement with the Orosz value. We therefore adopt $K_2 = 132.9 \pm 4.0$\,\kms\ from \citet{Orosz_2001} for all subsequent mass estimates.

\begin{figure}
    \includegraphics[width=1.0\linewidth, bb=0 0 990 700, clip=]{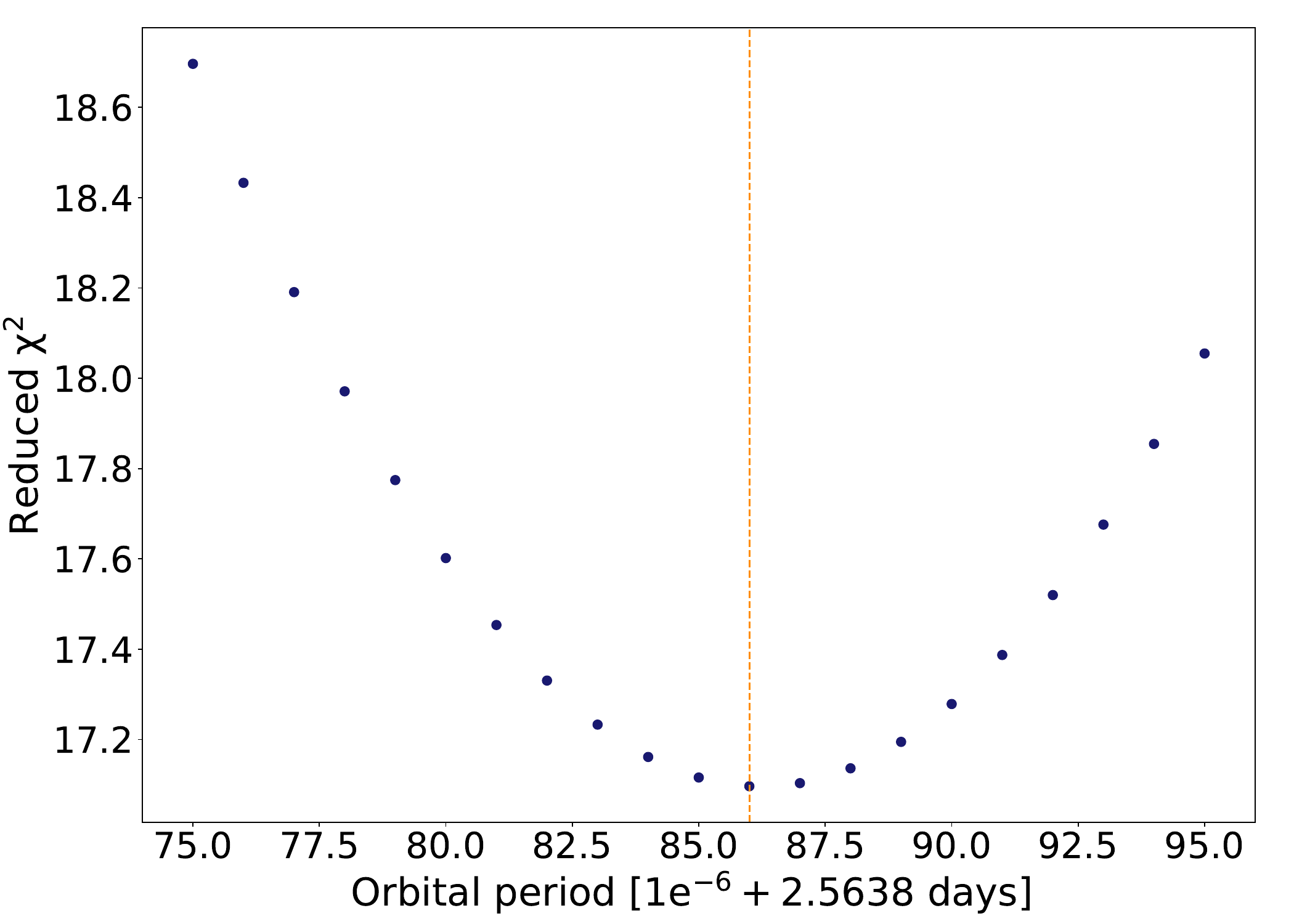}
    \caption{Orbital period optimization based on reduced chi-square.}
    \label{fig:Chi_squared}
\end{figure}

\subsection{Refinement of the Orbital Period}
\label{subsec:optim}

The orbital periods derived independently from the spectroscopic and photometric datasets are close but not identical. The discrepancies arise primarily from the nature of the data: the spectroscopic observations are sporadic and span a relatively short time baseline, whereas the photometric observations cover a much longer temporal range but exhibit changes in the light-curve morphology between epochs, with non-sinusoidal modulations at times. 

To refine the orbital period and reconcile the two determinations, we extended the time baseline by adopting the initial epoch ($T_{0}$) reported by \citet{Thorstensen_2017}. We recalculated both radial-velocity and photometric phase curves from that reference point, encompassing E$~\approx3800$ orbital cycles. We then explored a finely sampled range of trial periods between 2.563875 and 2.563895\,d, which brackets the values obtained from spectroscopy and photometry. For each trial period, we computed the radial-velocity fits and evaluated the corresponding root-mean-square (RMS) residuals and reduced $\chi^{2}$ statistics. 

A clear and well-defined minimum in $\chi^{2}_{\nu}$ was obtained for $P_{\mathrm{orb}} = 2.563886(3)$\,d, as shown in Figure~\ref{fig:Chi_squared}, improving the precision of the orbital period by one additional decimal digit with respect to the value derived by \citet{Thorstensen_2017}. Using this refined period and the adopted $T_{0}$, we define the updated ephemeris as

\begin{equation}
T_{\mathrm{MID}}(\mathrm{BJD}) = 2450709.76 + 2.563886(3)E.
\end{equation}

Figures~\ref{fig:tess_aavso} and~\ref{fig:vel_sec} were phased using this refined orbital period, which provides an optimal alignment between the spectroscopic and photometric datasets.

\begin{figure}
    \centering
    \includegraphics[width=1.0\linewidth]{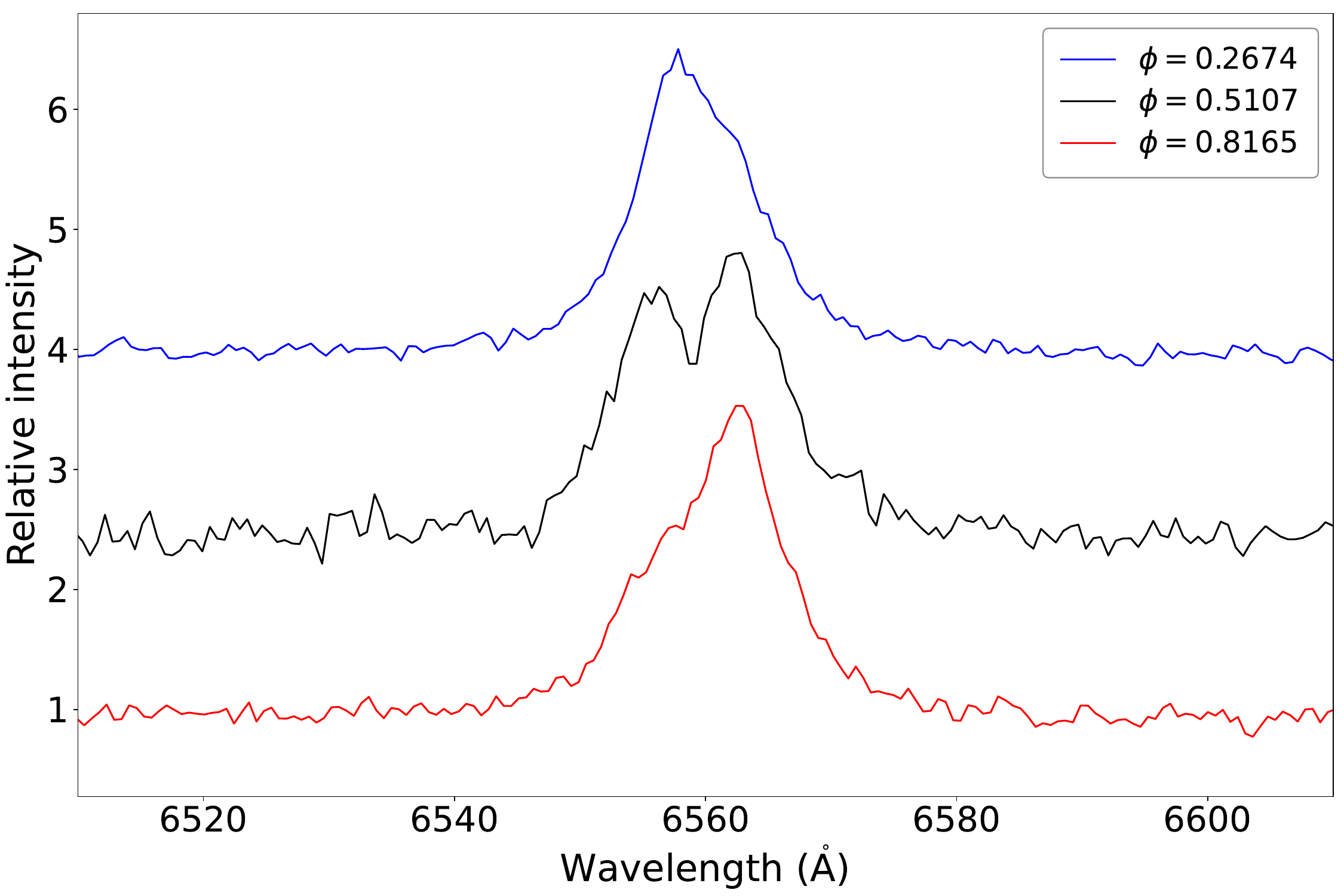}
    \caption{H$\alpha$ profiles in three different phases. In spectra with $\phi\sim 0.25$ (blue) and $\phi\sim 0.8$ (red) a single-peak profile is observed, loaded towards the left or right side of the profile centroid depending on the phase. In $\phi\sim 0.5$ (black), a double-peak profile is present. More details are shown in the text.}
    \label{fig:halphaprofile}
\end{figure}

\section{\texorpdfstring{H$\alpha$}{TEXT} EMISSION PROFILE AND RADIAL VELOCITY DECOMPOSITION}
\label{sec:Halpha}

The H$\alpha$ emission line of \vcas presents a notable morphology: the phase-averaged profile is largely single-peaked, with the centroid systematically blueshifted relative to the systemic velocity. This is unexpected for a disc-accreting binary at a high inclination $i \approx 70^{\circ}$ (see Section~\ref{subsec:MassFuncWD}), where the projected rotation of the disc would normally produce a clearly double-peaked line with peak separation of several hundred \kms. Instead, the profile width and shape are consistent with emission dominated by a large, optically thick accretion disc whose outer, optically thin rim contributes most of the line flux and whose azimuthal brightness asymmetry shifts the apparent centroid. At certain orbital phases, a secondary shoulder or asymmetry becomes evident on the red wing of the profile, and individual spectra show a resolved double-peaked structure whose separation and relative strength vary with phase, as shown in Figure \ref{fig:halphaprofile} for three different phases. This phase-dependent asymmetry, rather than a persistent double peak, motivates the decomposition described below.

To characterize the evolution of this asymmetry with orbital phase, we parameterized the H$\alpha$ profile at each epoch with two Gaussian functions. We emphasize that this decomposition is a mathematical description of a continuously varying, asymmetric disc profile — the two Gaussians track the centroid and relative strength of the blue and red sides of the emission feature, and do not in themselves imply two physically distinct emission regions. The physical interpretation is addressed in Section~\ref{sec:tomo} and~\ref{sec:discuss}.

\begin{figure}
    \centering
    \includegraphics[width=1.0\linewidth]{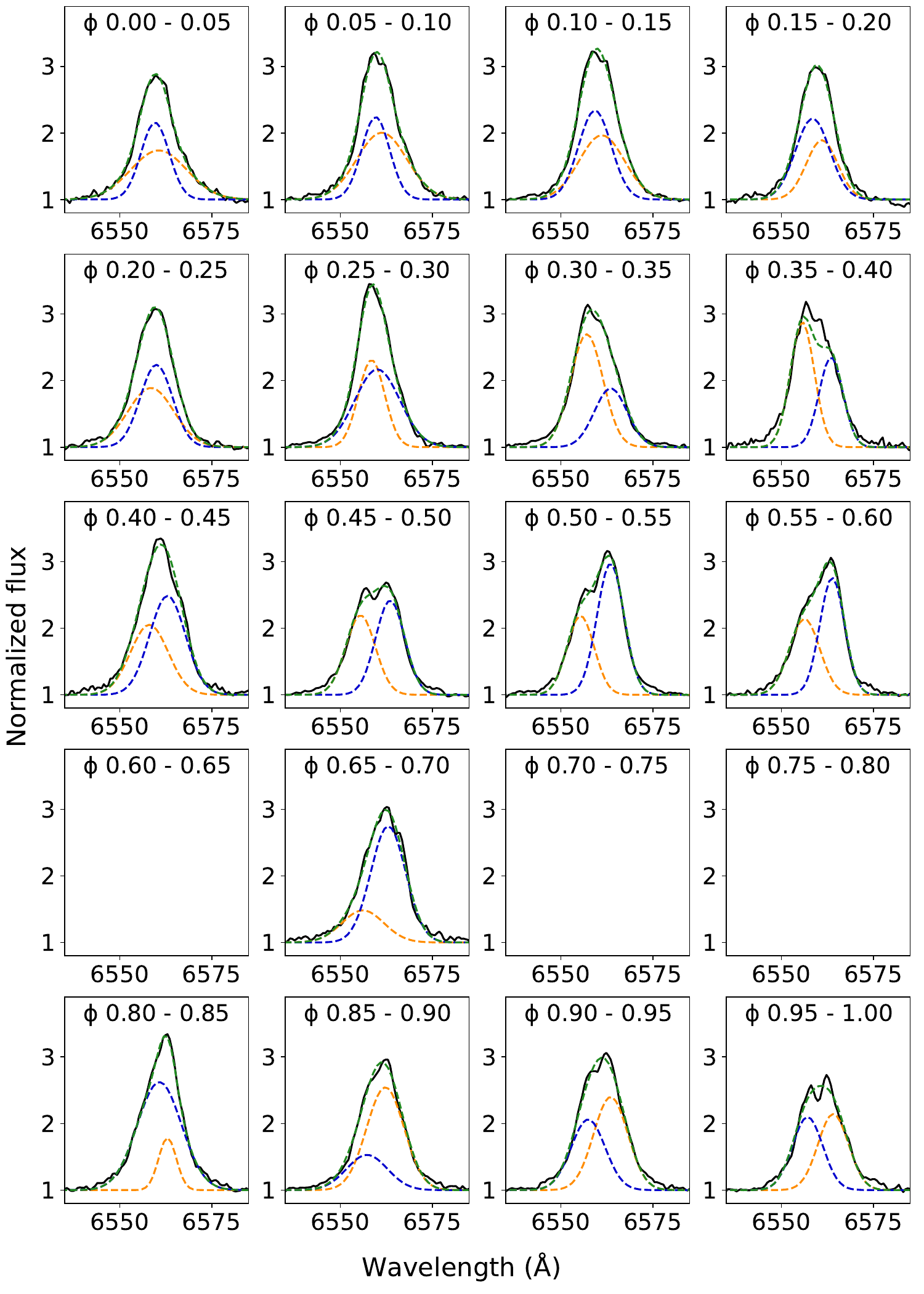}
    \caption{Two-Gaussian parameterization of the H$\alpha$ emission profile for each of the 17 phase-binned averaged spectra. The solid black line is the averaged spectrum; the orange and blue dotted lines are the individual Gaussian fits (components A and B respectively), and the green dotted line is their sum. The decomposition tracks the phase-dependent asymmetry of a single, broad disc emission feature rather than resolving two physically distinct emitters.}
    \label{fig:collagecomponentes}
\end{figure}

We measured the RVs of the two Gaussian components using the {\sc splot} task in the {\sc iraf} package with the interactive \emph{deblending} option \citep{Tody_1986,Tody_1993}. For each spectrum, we fit the H$\alpha$ emission profile with the minimum number of Gaussians needed to reproduce the observed asymmetric structure (as shown in Figure~\ref{fig:collagecomponentes}). The deblender returns, per component, the best–fit line center, integrated flux, equivalent width, core intensity, and Gaussian {\sc fwhm} together with the local continuum estimate.\footnote{See the {\sc splot} documentation for the deblend mode and returned parameters \url{https://iraf.readthedocs.io/en/latest/tasks/ctio/splot.html.}} These measurements were recorded in a log file with two rows per spectrum (one per Gaussian component). For each exposure, we associated the pair of RVs with the corresponding HJD and orbital phase from our observing log. The working matrix, therefore, has one row per spectrum with $\{\mathrm{HJD},\ \phi,\ v_1,\ v_2\}$ plus the auxiliary Gaussian parameters for both components (flux, EW, core, FWHM). Because the deblender does not label which Gaussian belongs to which physical component, the index $(1,2)$ attached to a given exposure may be swapped between components from one spectrum to the next. To disentangle the tracks, we modeled the two RV curves as circular–orbit sinusoids that share a common systemic velocity $\gamma$.

This choice enforces that the two curves cross at an unique systemic velocity, as expected for the two components measured in the same rest frame. We ignore gross outliers with $|v|>400$\,km\,s$^{-1}$ (typical mis-assignments or poor fits). We test below this unique systemic velocity assumption.

\begin{figure}
    \centering
     \includegraphics[width=1.0\linewidth]{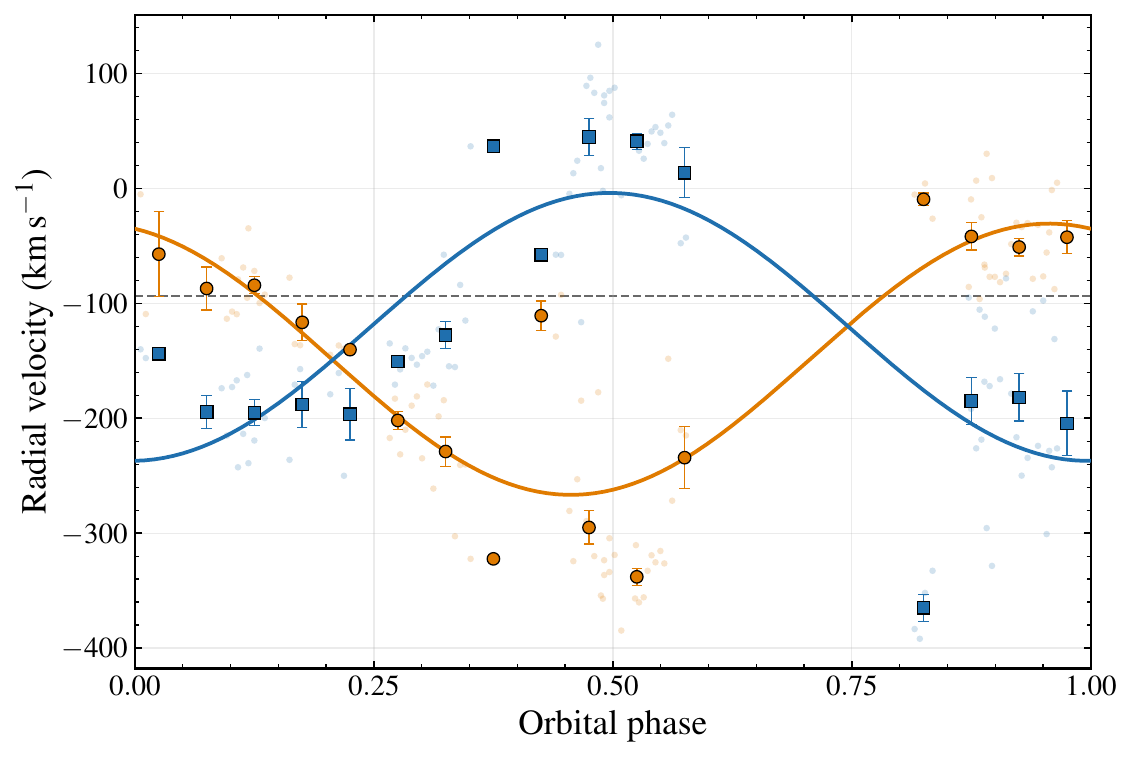}
    \caption{Radial velocities of components A (orange) and B (blue). Circular and square points represent the velocities measured from the phase-binned spectra, with best-fit semi-amplitudes $K_{\rm A} = 118$ km s$^{-1}$ and $K_{\rm B} = 117$ km s$^{-1}$, and systemic velocities $\gamma_{\rm A} = -149$ km s$^{-1}$ and $\gamma_{\rm B} = -120$ km s$^{-1}$, respectively. The faded points show the velocities measured from the individual spectra. The solid lines display the radial-velocity curves calculated from the phase-binned velocities only. The dashed line marks the adopted systemic velocity of the system, $\gamma = -93.5$ km s$^{-1}$.}
    \label{fig:vel_compfija}
\end{figure}

Calculating each sinusoid with a linear form:
\begin{equation}
v(\phi) = \gamma + A\sin(2\pi\phi) + B\cos(2\pi\phi),
\end{equation}
we solve for $(\gamma,A_A,B_A,A_B,B_B)$ by linear least squares while \emph{constraining} the per-spectrum assignment: exactly one of the $\{v_1,v_2\}$ velocities must be placed on fit $A$ and the other on fit $B$. We initialize with a trivial assignment fit ($v_1\!\to\!A$, $v_2\!\to\!B$) for both curves, and then iterate a deterministic ``swap'' step; for each exposure, we compute the total squared residual, keeping the current assignment or swapping the two measurements, and adopt the lower residual choice. We then refit and repeat until the assignments stabilize (typically in ${\sim}50$ iterations).

Finally, we recover $(K,\phi_0)$ from $(A,B)$ via
\begin{equation}
K=\sqrt{A^2+B^2},\qquad
\phi_0=\frac{1}{2\pi}\,\mathrm{atan2}(-B,\,A)\ \ (\mathrm{mod}\ 1),
\end{equation}
and obtain $\gamma$, $K$, and $\phi_0$ for both curves together with the per–point residuals. This approach is equivalent to a constrained two–cluster assignment in phase–velocity space, with the cluster centers restricted to sinusoidal orbits appropriate for circular spectroscopic binaries \citep[e.g.,][]{Hilditch_2001}. These two components, after the separation, are presented in the RV(A, B) vs. orbital phase Figure\,\ref{fig:vel_compfija} (faded orange and blue dots).

Due to the large dispersion in the RVs measured for each spectrum, we divided the spectra into 20 phase bins. We computed an average spectrum for each bin to improve the signal-to-noise ratio. This is depicted in Figure \ref{fig:collagecomponentes}, where we show the averaged spectrum as a solid black line; the orange and blue dotted lines correspond to Gaussian fits to the A and B components, respectively, and the green dotted line shows the sum of the components. In three cases, we did not obtain spectra within the bin range, so we ended up with only 17 mean spectra. We performed a two--component fit on these averaged spectra, from which we measured the radial velocities using the $\lambda$ corresponding to the center of the Gaussians. The velocity assignment for each component was performed in a manner similar to what was described above, but with one distinction - the systemic velocity was held free. There is a slight inconsistency between measurements of individual and binned spectra, because the number of spectra in each bin was not the same. In Figure~\ref{fig:vel_compfija} we show the radial-velocity curves which were constructed using just the velocities of the bins (circular and square dots)

The fitted parameters, derived from the 96 quality-selected spectra, already point compellingly toward a disc rather than a two-body interpretation. Most striking is the near-equality of the two amplitudes, $K_A \approx 118$\,\kms\ and $K_B \approx 117$\,\kms, combined with a phase separation $\Delta\phi = \phi_A - \phi_B \approx 0.46 \approx 0.5$. Two components almost equal in amplitude and almost exactly anti-phase is the hallmark signature of two Gaussians being forced to describe a single broad feature whose centroid shifts sinusoidally — when one wing moves blueward, the other moves redward by the same amount. Furthermore, neither systemic velocity (whether shared at $\gamma = -136$\,\kms, or individual at $\gamma_A = -148.6$ and $\gamma_B = -120.5$\,\kms) matches the true systemic velocity $\gamma = -93.5$\,\kms\ determined independently from the absorption-line fit. No emitter co-rotating in the binary frame can produce a systemic velocity offset of this magnitude. Both components are most naturally understood as the blue and red portions of a single, azimuthally asymmetric disc emission profile whose centroid and asymmetry vary continuously with orbital phase. The Doppler tomography in Section \ref{sec:tomo} confirms such interpretation.

\section{SED AND ESTIMATES OF THE STELLAR COMPONENT'S MASSES}
\label{seccion:DonorParam}

\subsection{Donor star mass}\label{subsec:donormass}

\vcas is a long-period CV whose binary parameters and component contributions were studied in detail by \citet{Orosz_2001}. Based on their spectroscopic analysis, the disc contributes about half of the flux in the optical domain; hence, the donor star, classified as a K4--K5\, IV star, is estimated to contribute approximately 50 percent of the flux.

\begin{figure}
\includegraphics[width=82.5mm, bb=0 0 630 330, clip=]{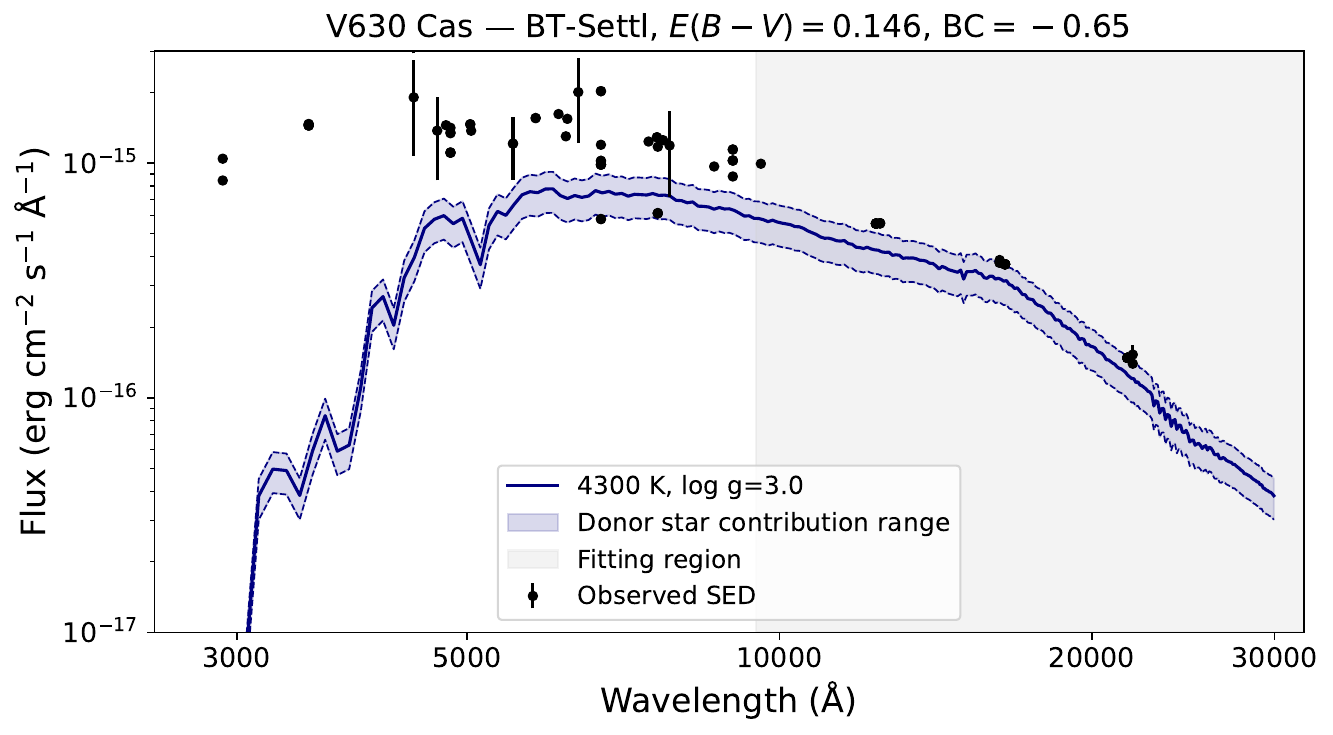}
\caption{Spectral energy distribution of V630~Cas (filled circles), de-reddened with $E(B-V)=0.146$. The navy curve is the BT-NextGen model at $T_{\rm eff}=4300$\,K, $\log g=3.0$, scaled to the NIR ($\lambda>9500$\,\AA) data; the shaded band shows the donor star contribution range for the Bailer-Jones distance interval $D=2216$--$2710$\,pc.}
\label{fig:SEDVCas}
\end{figure}

To constrain the donor star's parameters, we performed a grid-based fitting procedure targeting the infrared portion of the SED, where the donor dominates the flux. Before presenting our fit, we note an important caveat affecting how it should be read. Because the donor fills its Roche lobe, it is tidally distorted into a teardrop shape elongated toward the WD, and its projected area --- and hence its NIR flux at fixed $T_{\rm eff}$\ -- varies with orbital phase, being smallest near conjunction ($\phi = 0, 0.5$, when the elongated axis points along the line of sight) and largest near quadrature ($\phi = 0.25, 0.75$, when it is viewed side-on). For the closely analogous donor geometry in SDSS\,J085210.48+783246.6, we find this effect increases the projected flux by a factor of $\simeq 1.25$ from conjunction to quadrature \citep{Tovmassian_2026}. The archival $J, H, K$ photometry used below is a single-epoch measurement of unknown orbital phase and is therefore very unlikely to have been taken at exact conjunction; it should on average sit above a conjunction-phase (minimum-area) model at the same $T_{\rm eff}$. This is the origin of the offset between the data points and the illustrative curve in Figure~\ref{fig:SEDVCas}, which is scaled to the reddest ($\lambda > 9500$\,\AA) photometry and corresponds to a spherical approximation of the donor at its minimum projected area, as stated in the figure caption.

The photometry was also corrected for interstellar extinction, using $E(B-V) = 0.146 \pm 0.03$\,mag taken from the \citet{Green_2019} 3D dust map along the line of sight to \vcas, and adopting the standard $R_V = 3.1$ reddening law. Propagating this $\pm 0.03$\,mag uncertainty through the same Monte Carlo grid used below for the donor mass and radius changes $R_2$ by $< 0.02\ $,\rsun --- subdominant to the distance- and temperature-driven terms, and already included in the 68\,\%/95\,\%\ envelopes quoted below.

\begin{figure}
\includegraphics[width=\columnwidth]{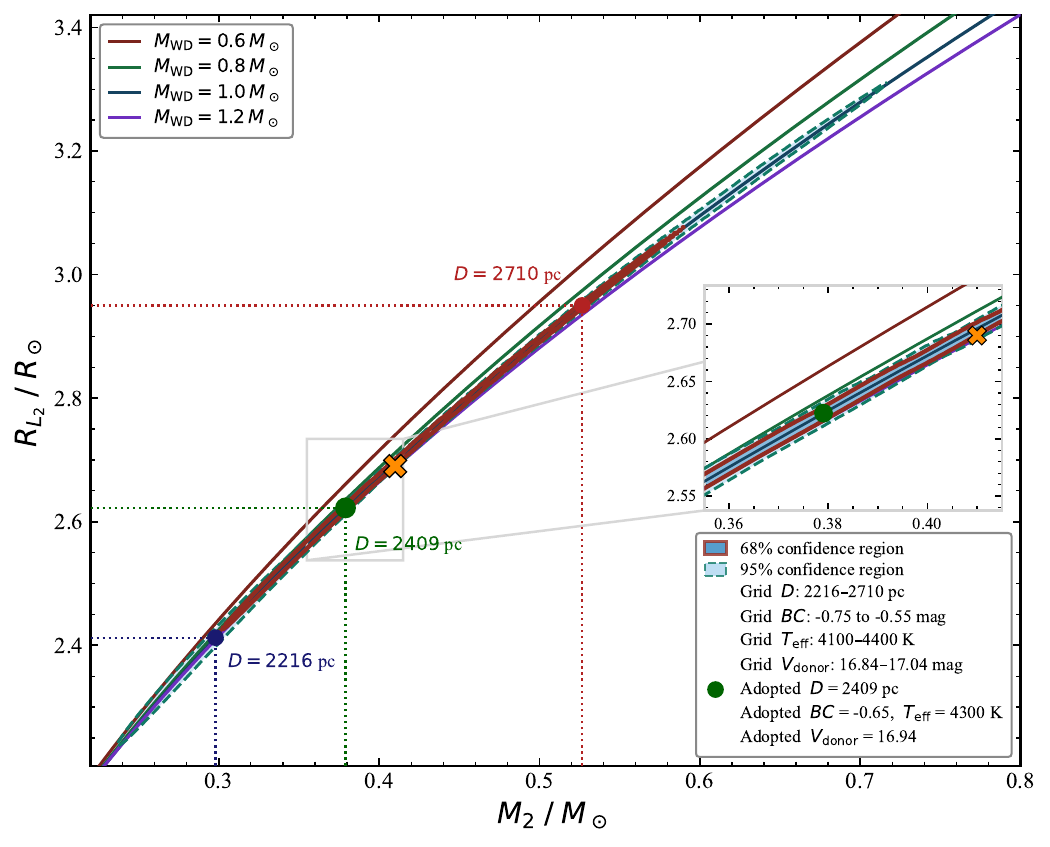}
\caption{Donor star Roche-lobe diagnostic diagram. Curves show $R_{L_2}$ vs.\ $M_2$ for $M_{\rm WD}=0.6$--$1.2$ \msun. The shaded regions represent the 68\,\% (dark blue) and 95\,\% (light blue) confidence intervals on the donor $(M_2,\,R_2)$, derived by propagating all four uncertain inputs simultaneously through the Stefan-Boltzmann radius marked as "Grids" in the bottom-right legend. Dotted crosshairs mark the solutions at the Bailer-Jones distance bounds with all other parameters at their adopted values ($BC=-0.65$, $T_{\rm eff}=4300$\,K, $V_{\rm donor}=16.94$). The orange cross marks the evolutionary model prediction; see Section\,\ref{sec:evolution}. The donor solution itself is essentially independent of $M_{\rm WD}$ (Section~\ref{subsec:MassFuncWD}); the self-consistent white-dwarf mass adopted below, $M_1\simeq1.27$\,\msun, lies slightly beyond the uppermost ($1.2$\,\msun) Roche-lobe curve shown here.}
\label{fig:Diagnostic}
\end{figure}

We then modeled the infrared SED to further constrain the donor star's physical parameters, where its contribution to the observed flux is expected to be dominant. Synthetic BT-NextGen spectra \citep{Allard_2012} with parameters appropriate for a K4--K5\,IV donor were fitted to the observed $J$, $H$, and $K$ magnitudes (Figure~\ref{fig:SEDVCas}) using $\chi^2$ minimization. This fitting approach leverages the Rayleigh-Jeans tail of the spectrum, which is less affected by contamination from the accretion disk and thus provides a robust constraint on the donor's physical size and the system's distance. The best-fitting solutions clustered around $\teff = 4100$--$4400$\,K, up to $R_2 \approx 3.0$\,\rsun, within $D = 2409^{+302}_{-192}$\,pc, in agreement with the  \textsc{Gaia}-derived distance \citep{Bailer-Jones_2021Data}. The solution presented in  Figure~\ref{fig:SEDVCas} involves a synthetic spectrum with $\teff = 4300$, $\log g = 4.0$. The best-fit stellar radius is $\geq 2.28$\,\rsun. The synthetic photometry of the scaled donor yielded a $V$-band magnitude of $V_{\rm donor} = 16.94$, which was subsequently used to place \vcas on the diagnostic diagram (Figure~\ref{fig:Diagnostic}).

Correcting the flux normalization for the $\simeq 1.25$ area enhancement discussed above raises the inferred donor radius by a factor $\sqrt{1.25} \simeq 1.12$, i.e.\ from $R_2 \simeq 2.62$\,\rsun~to $R_2 \simeq 2.93$\,\rsun -- consistent with, rather than in tension with, the upper end of the SED-fit radius range already quoted above (``up to $R_2 \approx 3.0$\,\rsun''). This pushes the donor mass modestly higher, not lower, reinforcing rather than undermining the qualitative conclusion of this section that the donor is significantly more massive than found by \citet{Orosz_2001}. A fully self-consistent treatment would require matching each photometric epoch to the Roche-lobe projected area at its specific orbital phase, which is not possible with the unphased archival photometry available to us; we flag this as a systematic in the same direction as, and comparable in size to, the distance uncertainty already propagated into $R_2$ and $M_2$ (Section~\ref{subsec:MassFuncWD}).

To estimate the mass of the donor star, we use the $M_2/$\msun~versus $R_2/$\rsun~diagnostic diagram adapted from \citet{Tovmassian_2025} (Figure~\ref{fig:Diagnostic}).  Rather than propagating only the distance uncertainty, we evaluate the Stefan-Boltzmann radius on a four-dimensional grid spanning all uncertain inputs: distance $D = 2216$--$2710$\,pc \citep{Bailer-Jones_2021Data}, bolometric correction $BC = -0.55$ to $-0.75$\,mag \citep{Eker_2020}, effective temperature $T_{\rm eff} = 4100$--$4400$\,K, and donor $V$-band magnitude $V_{\rm donor} \pm 0.10$\,mag.  Each grid point is projected onto the $M_{\rm WD} = 1.0$ \msun~Roche-lobe curve, and a kernel-density estimate of the resulting cloud defines the 68\,\% and 95\,\% confidence regions shown in Figure~\ref{fig:Diagnostic}.  For the adopted values ($D = 2409$\,pc, $BC = -0.65$, $T_{\rm eff} = 4300$\,K, $V_{\rm donor} = 16.94$) we obtain $M_2 = 0.38$ \msun~and $R_2 = 2.62$ \rsun.  The 68\,\% region spans $M_2 = 0.30$--$0.59$\,\msun\ and $R_2 = 2.43$--$3.08$\,\rsun; the wider 95\,\% region ($M_2 = 0.24$--$0.71$\,\msun) likewise remains within this limit and captures the full envelope when all four inputs are simultaneously pushed to their extremes.  This is in contrast with the estimate of \citet{Orosz_2001} of $M_2 < 0.2$ \msun.

\subsection{Mass function, self-consistent inclination, and WD mass}
\label{subsec:MassFuncWD}

With the adopted $K_2 = 132.9 \pm 4.0$\,\kms\ and orbital period $P_{\rm orb} = 2.563886$\,d, the mass function of the secondary star is

\begin{equation*}
  f(M_1) = \frac{P_{\rm orb}\,K_2^3}{2\pi G}= \frac{M_1^3\,\sin^3 i}{(M_1 + M_2)^2} = 0.62 \pm 0.06\,\text{M}_\odot,
\end{equation*}
in agreement with \citet{Orosz_2001}. For comparison, using our own lower $K_2 = 122.8$\,\kms\ gives $f(M_1) = 0.49$ \msun, illustrating the strong sensitivity of the mass function to the adopted semi-amplitude ($K_2^3$).

Because both the donor absorption ($K_2$) and disc emission ($K_1$) semi-amplitudes are measured, we can form the velocity sum $K_1 + K_2$, which constrains the total mass independently of the individual values. \citet{Orosz_2001} measured $K_1 = 39.1 \pm 4.9$\,\kms\ from H$\alpha$ emission (single-Gaussian fit). Despite the complexity of the H$\alpha$ profile discussed in Sections~\ref{sec:tomo} and~\ref{sec:discuss}, the sum $K_1 + K_2 \approx 172$\,\kms\ provides a robust constraint: the total minimum mass is well-constrained independently of the individual component amplitudes,

\begin{equation*}
  (M_1 + M_2)\sin^3 i = \frac{P_{\rm orb}(K_1 + K_2)^3}{2\pi G} \approx 1.35\,\text{M}_\odot.
\end{equation*}
Ellipsoidal models fitted by \citet{Orosz_2001} to their mean light curve give an inclination in the range $66.96^{\circ} < i < 78.08^{\circ}$, rather than a single value. Evaluating the above over this full range, together with their mass ratio $q = K_1/K_2 = 39.1/132.9 = 0.29$, gives $M_1 \approx 1.12$--$1.34$\,\msun\ and $M_2 \approx 0.33$--$0.39$\,\msun. This purely spectroscopic estimate, which predates the revised donor radius of Section~\ref{subsec:donormass}, is superseded below by the self-consistent solution; we retain it here only as an independent order-of-magnitude cross-check, to which it remains reasonably close. Even at the most conservative limit $i = 90^\circ$, the minimum WD mass from the mass function alone is $M_1 \geq f(M_1)(1+q)^2 = 0.62 \times 1.29^2 \approx 1.03$ \msun.

As anticipated in Section~\ref{subsec:photometry}, the inclination cannot simply be inherited unchanged from \citet{Orosz_2001}: their ellipsoidal fit was obtained for a donor of radius $R_2\simeq2.0$\,\rsun\ (implied by their own $M_1=0.98$\,\msun, $M_2=0.17$\,\msun), whereas the SED- and diagnostic-diagram-based analysis above gives a significantly larger donor, $R_2=2.62^{+0.46}_{-0.19}$\,\rsun. We therefore re-derive $i$ self-consistently, combining the mass function above with Kepler's third law and the Eggleton Roche-lobe approximation \citep{Eggleton_1983}, and requiring consistency with the observational constraint (Section~\ref{subsec:photometry}) that no eclipse is detected at any TESS epoch.

For a given inclination $i$ and mass ratio $q=M_2/M_1$, the mass function fixes the white-dwarf mass through
\begin{equation*}
M_1(i,q) = \frac{f(M_1)\,(1+q)^2}{\sin^3 i},
\end{equation*}
and Kepler's third law fixes the orbital separation through
\begin{equation*}
a(i,q) = \left[\frac{G\,M_1(i,q)\,(1+q)\,P_{\rm orb}^2}{4\pi^2}\right]^{1/3}.
\end{equation*}
Requiring the Roche-lobe radius, $R_{L2}=\mathrm{Eggleton}(q)\,a$, to equal the photometrically determined donor radius $R_2=2.62$\,\rsun\ (Section~\ref{subsec:donormass}) then yields a unique $q(i)$ for each trial inclination, and hence $M_1(i)$, $M_2(i)$, and $a(i)$. 
Table~\ref{tab:inclination_grid} lists the resulting solution over the $66.96^{\circ}$--$78.08^{\circ}$ range determined by \citet{Orosz_2001}'s ellipsoidal fit (90\% confidence); note that $M_2$ is essentially independent of $i$ ($0.381$--$0.382$\,\msun\ throughout), since it is fixed almost entirely by the photometric donor radius, while $M_1$ increases steeply as $i$ decreases, ranging from $1.17$\,\msun\ at the upper end of the Orosz range to $1.33$\,\msun\ at the lower end.

\begin{table}
\centering
\caption{Self-consistent system parameters as a function of adopted inclination, with $f(M)=0.62$ \msun~and donor radius $R_2=2.62$ \rsun.}
\label{tab:inclination_grid}
\begin{tabular}{ccccc}
\hline
$i$ (deg) & $q$ & $M_1$ (\msun) & $M_2$ (\msun) & $a$ (\rsun) \\
\hline
66.96 & 0.288 & 1.33 & 0.382 & 9.43 \\
68.00 & 0.292 & 1.31 & 0.382 & 9.39 \\
70.00 & 0.300 & 1.27 & 0.382 & 9.32 \\
72.00 & 0.308 & 1.24 & 0.381 & 9.26 \\
74.00 & 0.314 & 1.21 & 0.381 & 9.21 \\
76.00 & 0.320 & 1.19 & 0.381 & 9.16 \\
78.08 & 0.325 & 1.17 & 0.381 & 9.12 \\
\hline
\end{tabular}
\end{table}

1. Upper bound from the absence of an eclipse. At phase 0.0, the donor transits in front of the disc; the requirement that it not occult the disc places a purely geometric upper limit on $i$. For the conservative case in which only the compact central region (effectively a point source at the disc centre) must remain unocculted, the condition $a(i)\cos i > R_2$ evaluates to $i \lesssim 73.5\degr$ (point-source / disc-centre limit). This is already $\sim2\degr$ tighter than the limit implied by the original Orosz donor ($i\lesssim76\degr$ for $R_2=2.0$\,\rsun), a direct consequence of the larger donor radius found here. If instead the full spatial extent of the ``large, optically thick'' disc (Section~\ref{sec:Halpha}) is required to remain unocculted -- adopting a tidally truncated disc radius $R_{\rm disc}\simeq0.7\,R_{L,\rm WD}$, consistent with the truncation fraction used for SDSS\,J0852+7832 \citep{Tovmassian_2026} -- the bound tightens substantially further, to $i\lesssim52\degr$.

\begin{figure}
\centering
\includegraphics[width=\columnwidth]{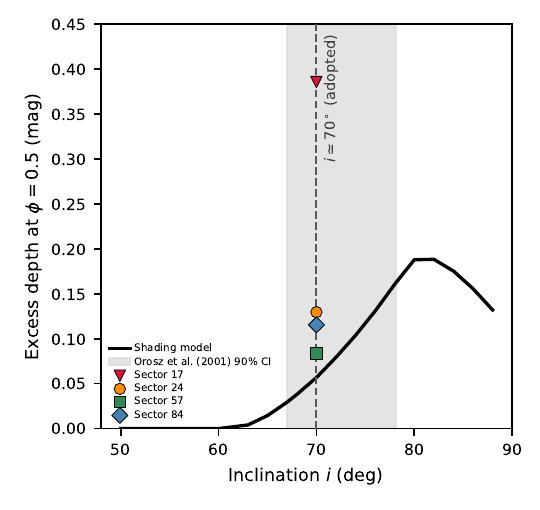}
\caption{Predicted excess depth of the phase-0.5 minimum relative to phase-0.0, from the flared-disc shading model described in the text, as a function of inclination (solid curve). The shaded band marks the $67^{\circ}$--$78^{\circ}$ range from the ellipsoidal light-curve fit of \citet{Orosz_2001}; the dashed line marks the adopted fiducial value $i=70^{\circ}$.} Coloured points show the observed excess in each TESS sector, plotted at $i=70^{\circ}$ for reference: Sectors 24, 57, and 84 lie within, or close to, the range spanned by the model over the Orosz confidence interval, while Sector 17 lies well above the maximum the steady-state model can produce at any inclination (see text).
\label{fig:ShadingModel}
\end{figure}

2. Lower bound from the disc-shading asymmetry. A softer, independent constraint comes from the observed excess depth of the minimum at phase 0.5 relative to phase 0.0 (the ellipsoidal asymmetry discussed in Section~\ref{subsec:photometry} and by \citet{Orosz_2001}), attributed to partial shading of the donor by the flared outer disc rim at superior conjunction. We model this with a standard flared-disc geometry, $H(R)=H_{\rm rim}(R/R_{\rm disc})^{1.125}$ with $H/R=0.13$ at the rim \citep{Tovmassian_2026}, and ray-trace the donor's visible hemisphere against the disc silhouette at phase 0.5 for trial inclinations (Figure~\ref{fig:ShadingModel}).  The predicted shading excess is small ($<0.03$~mag) below $i\simeq63$--$67\degr$ and rises to $\sim0.15$--$0.19$~mag by $i\simeq78$--$82\degr$. The observed excess in the three sectors that show a stable, single-peaked ellipsoidal profile (Sectors 24, 57, and 84; $0.08$--$0.13$~mag) requires $i\gtrsim65\degr$ to be produced at all by this mechanism.

3. Combined constraint. Both independent lines of evidence above are compatible with the inclination range determined directly from the ellipsoidal light-curve fit of \citet{Orosz_2001}, $66.96\degr < i < 78.08\degr$ (90\% confidence): the absence of an eclipse excludes only inclinations toward the upper end of this range ($i\gtrsim73.5\degr$ for the point-source criterion), while the disc-shading asymmetry favours inclinations toward its lower half ($i\gtrsim65\degr$). Rather than deriving an independent range, we therefore adopt the \citet{Orosz_2001} range directly, taking $i\simeq70\degr$ -- toward its lower half, where both constraints above are most comfortably satisfied -- as our fiducial value, for which the relations above give $M_1\simeq 1.27^{+0.10}_{-0.06}$\,\msun,\, $M_2\simeq0.38$\,\msun,\, $q\simeq 0.30$. This value of $M_1$ sits close to the middle of the $1.12$--$1.34$\,\msun\ range obtained above directly from the Orosz $K_1+K_2$ constraint over their full inclination range, providing an independent cross-check on the self-consistent solution. This revision raises the white dwarf mass relative to the lower end of that cross-check range: enforcing consistency with the non-detection of an eclipse does not alleviate the high-mass tension noted for this system; it sharpens it. We adopt $i\simeq70\degr$, $M_1\simeq1.27$\,\msun, $M_2\simeq0.38$\,\msun, and $q\simeq0.30$ throughout the remainder of this paper.

Sector 17 remains a notable outlier: its phase-0.5 minimum is $\sim0.39$~mag deeper than at phase 0.0, roughly twice the maximum excess ($\sim0.19$~mag) that the steady-state flared-disc model can produce at any inclination (Figure~\ref{fig:ShadingModel}). Combined with the $\delta\phi\simeq0.08$ phase displacement of that minimum noted in Section~\ref{subsec:photometry}, this is most naturally explained by a transient, non-axisymmetric enhancement of the disc rim height during that epoch -- consistent with the precessing/eccentric disc scenario already invoked to explain the sector-to-sector morphology changes -- rather than by a revision of the mean system inclination itself.

We emphasise that the inclination range derived above is a geometric bound, not a substitute for direct light-curve modelling. A definitive inclination and white-dwarf mass will ultimately require fitting the full TESS light curve with a code such as ELC, as originally used for this system by \citet{Orosz_2001}; the epoch-to-epoch variability of the disc geometry documented in Section~\ref{subsec:photometry}, however, means that such modelling must itself account for a time-variable disc, and it is beyond the scope of this study.

\begin{figure*}[t]
\setlength{\unitlength}{1mm}
\resizebox{11.2cm}{!}{
\begin{picture}(100,50)
\put (0,0){\includegraphics[width=35mm, bb=0 0 426 556, clip=]{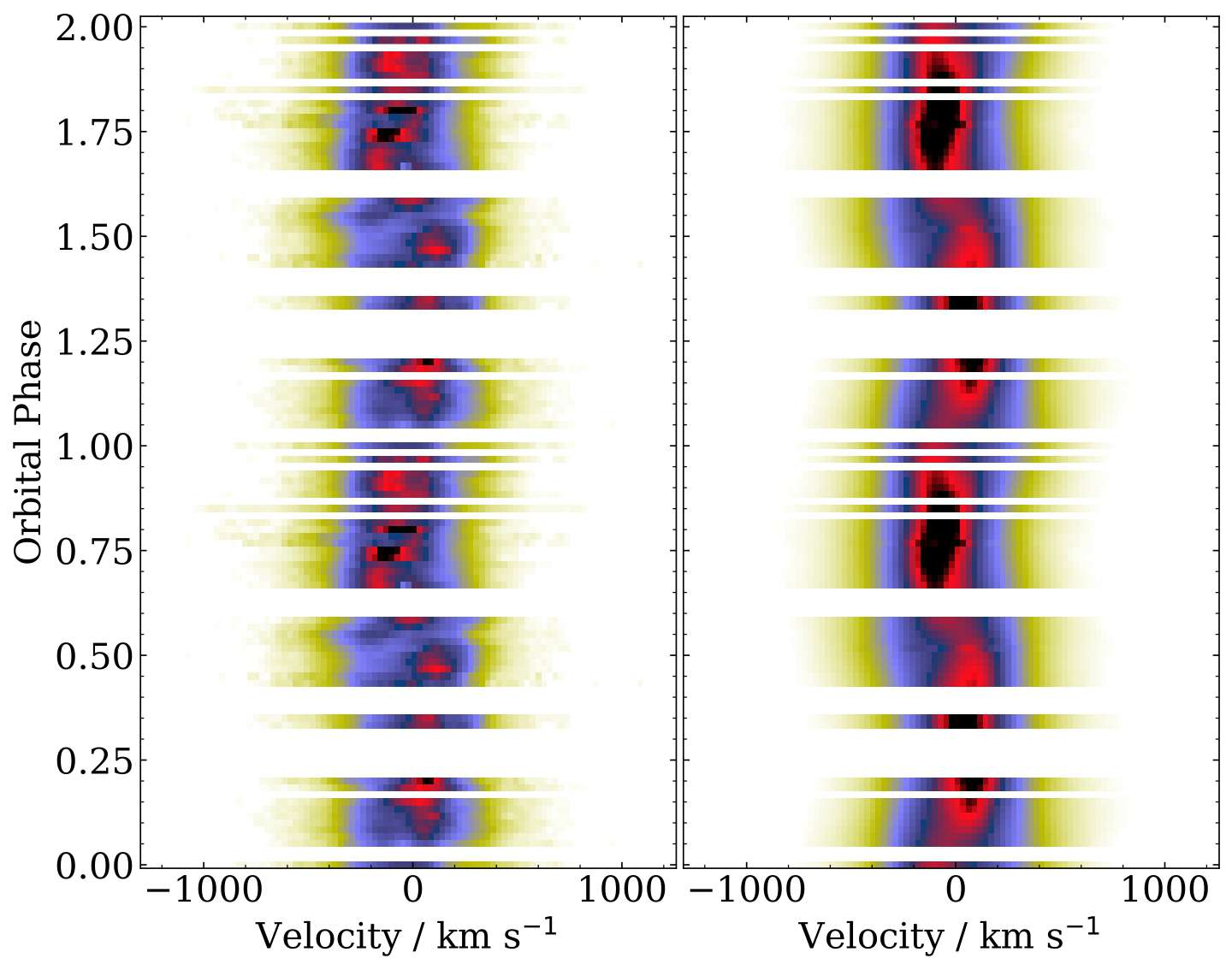}}
\put (31.4,0){\includegraphics[width=197mm, bb=0 0 2750 2597, clip=]{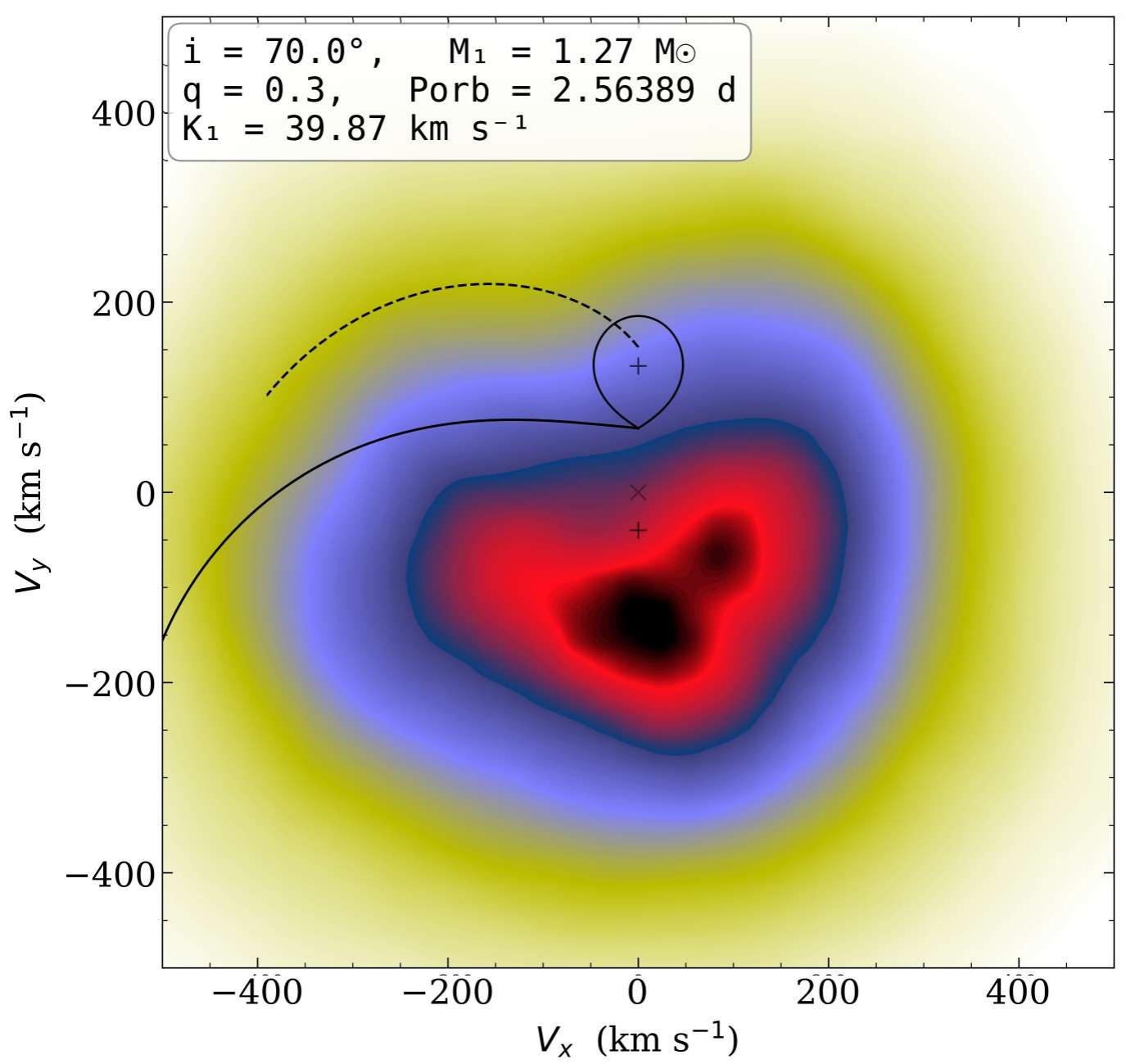}}
\put (79,0){\includegraphics[width=35mm, bb=0 0 426 556, clip=]{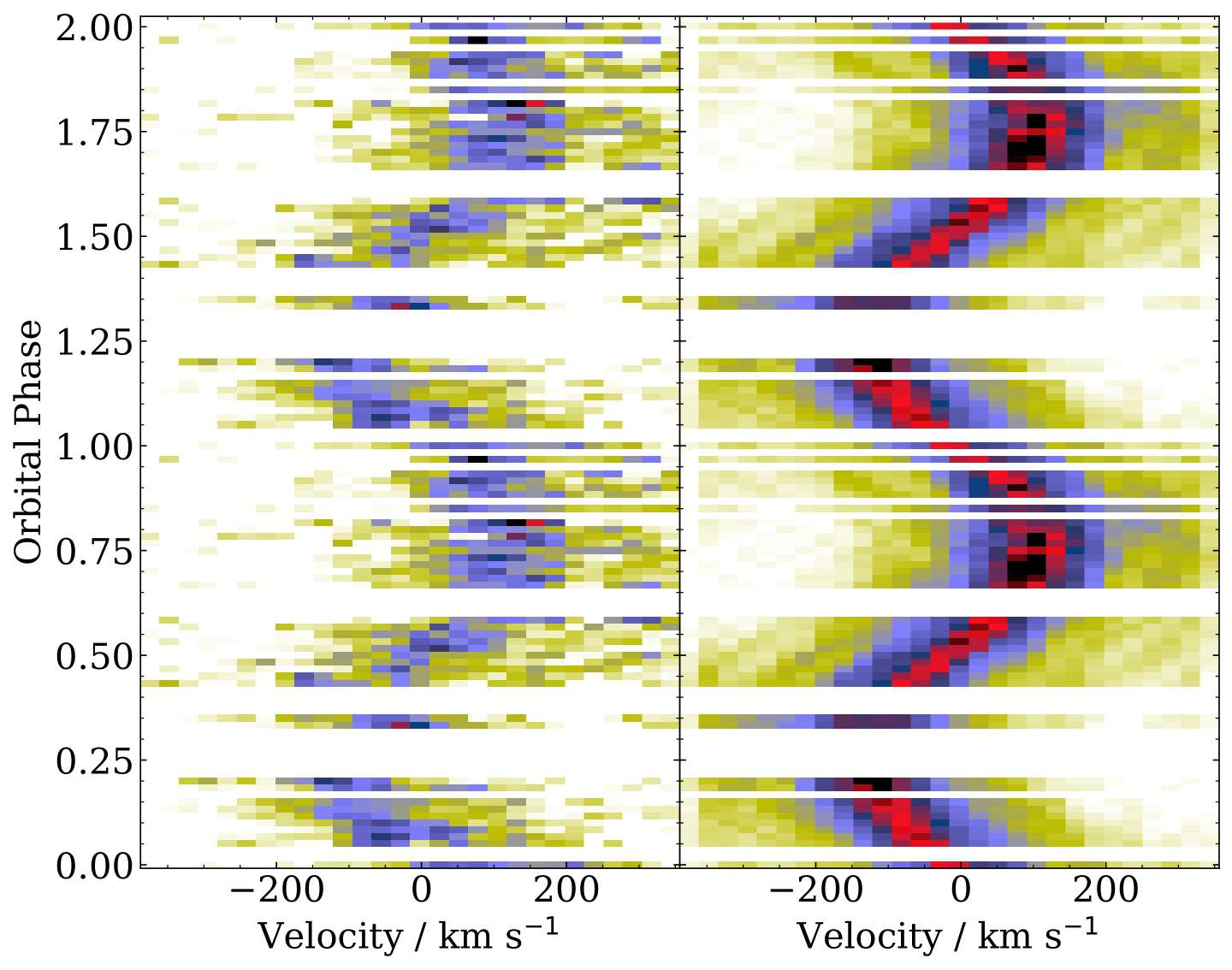}}
\put (110.5,0){\includegraphics[width=63.5mm, bb=0 0 2753 2602, clip=]{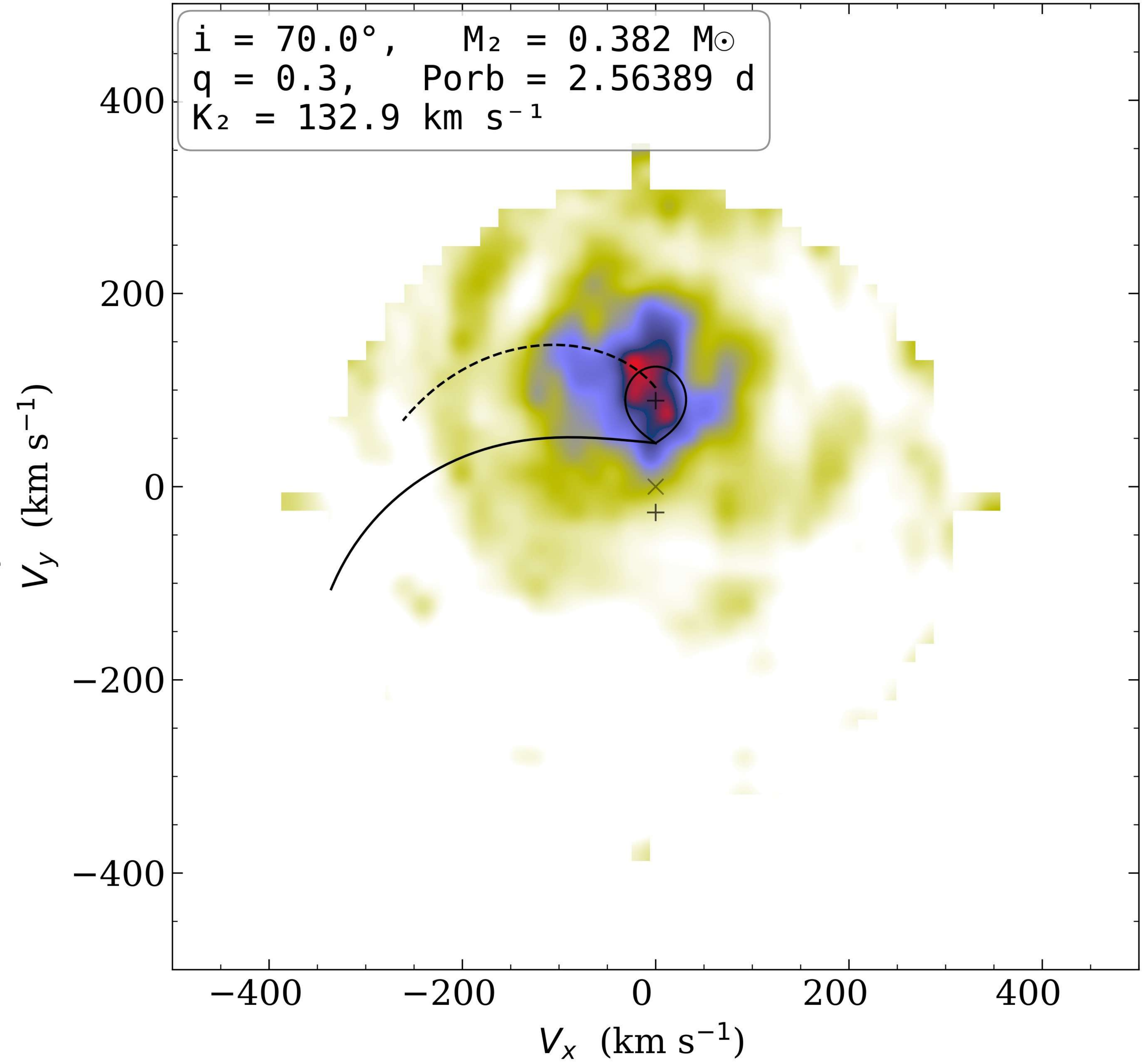}}
\end{picture}}
\caption{Trailed spectra and Doppler tomograms of \vcas\ in H$\alpha$ emission (left two panels) and Fe\,{\sc i} $\lambda \ 6494.98$ \AA~absorption (right two panels), constructed from all individual spectra using $i = 70^{\circ}$, $M_1 = 1.27$~\Msun, $M_2 = 0.38$~\Msun, $q = 0.30$, $K_1 = 39$~\kms, $K_2 = 132.9$~\kms and  $P_{\rm orb} = 2.563886$\,d. The Keplerian and ballistic trajectories of the gas stream are marked as the upper and lower curves, respectively. The H$\alpha$ tomogram shows emission distributed in a broad, azimuthally asymmetric pattern centered on the WD, with peak intensity toward blueshifted velocities; no compact hot-spot is detected at the expected disc--stream impact location. The Fe\,{\sc i} tomogram reveals the donor star absorption signal, confirming the donor detection independently of the radial-velocity curve.}
\label{fig:DopplerMap}
\end{figure*}

\subsection{Tomography}
\label{sec:tomo}

Doppler Tomography is a spectroscopy technique that uses phase-resolved line profiles to map the accretion flow in velocity space. A detailed formulation of this technique can be found in \citet{Marsh_1988} and \citet{Marsh_2005}. To obtain the tomography, we used the program {\sc pydoppler}\footnote{\url{https://github.com/Alymantara/pydoppler}} adapted by~\citet{Hernandez_2021} (and further improved by ourselves) from \citet{Spruit_1998}. Trailed spectra ({\sc trail}) were obtained, and Doppler tomography was constructed ({\sc stream}). Additionally, a reconstruction of the trailed spectra ({\sc reco}) was generated from the obtained Doppler tomography. Tomography was performed using the H$\alpha$ emission line of all individual spectra (Figure~\ref{fig:DopplerMap}, left two panels). The system parameters adopted are: inclination $i = 70^{\circ}$, mass ratio $q = M_2/M_1 = 0.30$, $M_1 = 1.27$ \Msun, $K_1 = 39$\,\kms\ (from the mass ratio) and \porb$=2.563886$~d.

The H$\alpha$ Doppler tomogram shows emission distributed in a broad region centered on the WD, with azimuthal asymmetry and peak intensity extending toward blueshifted velocities. Notably, no compact spot is present at the expected disc--stream impact location ($V_x > 0$, $V_y \sim +K_2$), and the phase of maximum emission is inconsistent with a classical hot-spot S-wave.  We interpret this morphology as emission from the optically thin outer rim of the accretion disc, whose azimuthal asymmetry produces the observed phase-dependent centroid variation. The epoch-dependent changes seen in the \textsc{TESS} light curves suggest the disc geometry (and hence the emission asymmetry pattern) varies over timescales of years, possibly reflecting a precessing eccentric disc.

Additionally, the Doppler tomogram of the Fe\,{\sc i} $\lambda6494.98$~\AA~absorption line was obtained (Figure~\ref{fig:DopplerMap}, right two panels). Unlike the case of H$\alpha$, not all individual spectra were used. As mentioned in Section \ref{sec:observations}, some spectra exhibit a low signal-to-noise ratio, so these spectra were excluded from the Doppler tomography analysis of this line. Ultimately, the 111 spectra with a S/N~$\geq 15$ were used. The Doppler map and trailed spectrum of the donor absorption line (inverted intensities) show the donor star absorption signal within the Roche lobe contour. The Tomography reveals the absorption signal of the donor star, confirming the detection of the donor independently of the radial-velocity curve.

\section{BINARY EVOLUTION MODELS}
\label{sec:evolution}

\begin{table}
\caption{Stellar and binary evolution parameters were adopted in MESA. The best-fit values are highlighted in bold.}
\label{Tab:Assumptions}
\centering
\setlength\tabcolsep{0.05pt} 
\renewcommand{\arraystretch}{1.2} 
\begin{tabular}{lc}
\hline
Parameter & Value \\
\hline
Initial post-CE orbital period &  $2,\,\ldots,\textbf{12.4},\,\ldots,15$~d \\
Initial post-CE WD mass &  $0.8,\,0.9,\,1.0,\,\textbf{1.1}$~\Msun \\
Initial post-CE companion mass &  $1.0,\,\textbf{1.1},\,1.2,\,1.3$~\Msun \\
Metallicity Z  &  $0.010,\,0.015,\,\textbf{0.020}$ \\
Criterion for stable H burning & \citet{Wolf_2013} \\
Criterion for stable He burning & \citet{Kato_2004} \\
Magnetic braking model  &  CARB \citep{Van+Ivanova_2019} \\
Consequential AML model  &  mass loss \citep{King_1995} \\
Wind prescription  &  \citet{Reimers_1975} \\
Reimers's parameter &  0.5 \\
Mixing length theory & \citet{Henyey_1965} \\
Mixing length ($H_p$) & $\textbf{1.5},\,2.0,\,2.5$ \\
\vspace{-0.15cm}
Extent of diffusive & \multirow{2}{*}{$0.016\,H_p$}\\
exponential core overshooting &  \\
\vspace{-0.15cm}
Criterion for stability & \multirow{2}{*}{Schwarzschild $(\nabla_{\rm ad} = \nabla_{\rm rad})$} \\
against convection & \\
\vspace{-0.15cm}
\multirow{2}{*}{Opacity table} &  \textbf{\citet{Ferguson_2005}}, \\
& \citet{Freedman_2008,Freedman_2014} \\
Nuclear network & \texttt{cno$\_$extras.net} \\
Atmosphere boundary conditions & Eddington T($\tau$) relation \\
\vspace{-0.15cm}
How opacities are calculated  & \multirow{2}{*}{varying, \textbf{iterated}} \\
throughout the atmosphere & \\
\hline
\end{tabular}
\end{table}

Using the parameters of \vcas as constraints, we can search for reasonable evolutionary sequences able to explain its properties. For that purpose, we used the version r15140 of the \textsc{MESA} code \citep[][]{Paxton_2011,Paxton_2013,Paxton_2015,Paxton_2018,Paxton_2019,Jermyn_2023} to run a grid of post-common-envelope (post-CE) binary models. After finding the best-fitting post-CE model with the \textsc{MESA} code, we searched for a reasonable pre-CE model by running a grid of models with the rapid BSE code following the approach of \citet{Belloni_2024b}, assuming an efficiency of ${\approx30}$\% for the CE evolution.

In what follows, we describe our assumptions for single--star and binary evolution in our grid of \textsc{MESA} models, which are summarized in Table~\ref{Tab:Assumptions}. Our approach follows closely that by \citet{Belloni+Schreiber_2023}\,\footnote{\href{https://zenodo.org/records/8279474}{https://zenodo.org/records/8279474}}, \citet{Tovmassian_2025}, and \citet{Tovmassian_2026}.

The \textsc{MESA} equation of state is a blend of the OPAL \citep{Rogers_2002}, SCVH \citep{Saumon_1995}, FreeEOS \citep{Irwin_2004}, HELM \citep{Timmes_2000}, PC \citep{Potekhin_2010} and Skye \citep{Jermyn_2021} equations of state. Nuclear reaction rates are a combination of rates from NACRE \citep{Angulo_1999}, JINA REACLIB \citep{Cyburt_2010}, plus additional tabulated weak reaction rates \citep{Fuller_1985,Oda_1994,Langanke_2000}. Screening is included via the prescription of \citet{Chugunov_2007}, and thermal neutrino loss rates are from \citet{Itoh_1996}. Electron conduction opacities are from \citet{Cassisi_2007} and radiative opacities are primarily from OPAL \citep{Iglesias_1993,Iglesias_1996}, with the high-temperature Compton-scattering dominated regime calculated using the equations of \citet{Buchler_1976}.

We varied the metallicity by adopting the values ${Z=0.01}$, $0.015$, and $0.02$, consistent with a solar-like metallicity. We assumed the grey Eddington T($\tau$) relation to calculate the outer boundary conditions of the atmosphere. We adopted either a uniform opacity that is iterated to be consistent with the final surface temperature and pressure at the base of the atmosphere, or a varying opacity consistent with the local temperature and pressure throughout the atmosphere. For low-temperature radiative opacities, we adopted those from either \citet{Ferguson_2005} or \citet[][]{Freedman_2008,Freedman_2014}. We further used the nuclear network \texttt{cno$\_$extras.net}, which includes the nuclear reactions of both the proton-proton and the carbon-nitrogen-oxygen-hydrogen burning cycles.

\begin{table*}
\centering
\caption{Evolution of a zero-age main-sequence binary towards the present-day properties of \vcas. The row in which the binary has the present-day properties of \vcas is highlighted in boldface. Pre- and post-CE evolutions were calculated with BSE, and CE evolution with MESA.}
\label{Tab:FormationChannel}
\begin{tabular}{r c c c c c c r l}
\hline
 Time  &   $M_1$        &   $M_2$     & $T_{\rm eff,2}$  &    $R_2$     & Type$_1$    & Type$_2$ & $P_{\rm orb}$ & Event\\
 (Myr) &   (M$_\odot$)  & (M$_\odot$) &    (K)           &  (R$_\odot$) &             &          &  (d)          &      \\
\hline
   0.0000  &  5.812  &  1.045 &  5707 & 1.062  & MS     & MS  & 6050.0000  &  Initial zero-age MS binary \\
  73.6108  &  5.796  &  1.045 &  5735 & 1.069  & SG     & MS  & 6078.9766  &  Change in primary type \\
  73.8851  &  5.796  &  1.045 &  5735 & 1.069  & FGB    & MS  & 6079.7343  &  Change in primary type \\
  74.0109  &  5.795  &  1.045 &  5736 & 1.069  & CHeB   & MS  & 6081.8567  &  Change in primary type \\
  83.9613  &  5.712  &  1.045 &  5746 & 1.071  & E-AGB  & MS  & 6230.2196  &  Change in primary type \\
  84.4566  &  5.670  &  1.045 &  5747 & 1.071  & TP-AGB & MS  & 6297.7209  &  Change in primary type \\
  84.9241  &  3.179  &  1.100 &  5748 & 1.071  & TP-AGB & MS  & 5204.0320  &  Begin of RLOF (primary is the donor) \\
  84.9241  &  3.179  &  1.100 &  5748 & 1.071  & TP-AGB & MS  & 5204.0320  &  CE evolution \\
  84.9241  &  1.100  &  1.100 &  5748 & 1.071  & WD     & MS  &   12.4000  &  End of RLOF \\
6036.0579  &  1.100  &  1.098 &  5751 & 1.478  & WD     & SG  &   12.4228  &  Change in secondary type \\
7998.7249  &  1.100  &  1.096 &  4828 & 2.265  & WD     & FGB &   12.3069  &  Change in secondary type \\
8248.1231  &  1.100  &  1.096 &  4579 & 3.131  & WD     & FGB &    1.8808  &  Begin of RLOF (secondary is the donor) \\
\textbf{8252.5410}  &  \textbf{1.274}  &  \textbf{0.377} & \textbf{4288} & \textbf{2.607} & \textbf{WD} & \textbf{FGB} & \textbf{2.5639} & \textbf{Binary looks like \vcas} \\
8485.6235  &  1.274  &  0.187 &  5530 & 1.639  & WD     & Proto-WD &  1.8738  &  End of RLOF \\
\hline
\end{tabular}
\end{table*}

\begin{figure*}
\centering
\includegraphics[width=0.35\linewidth]{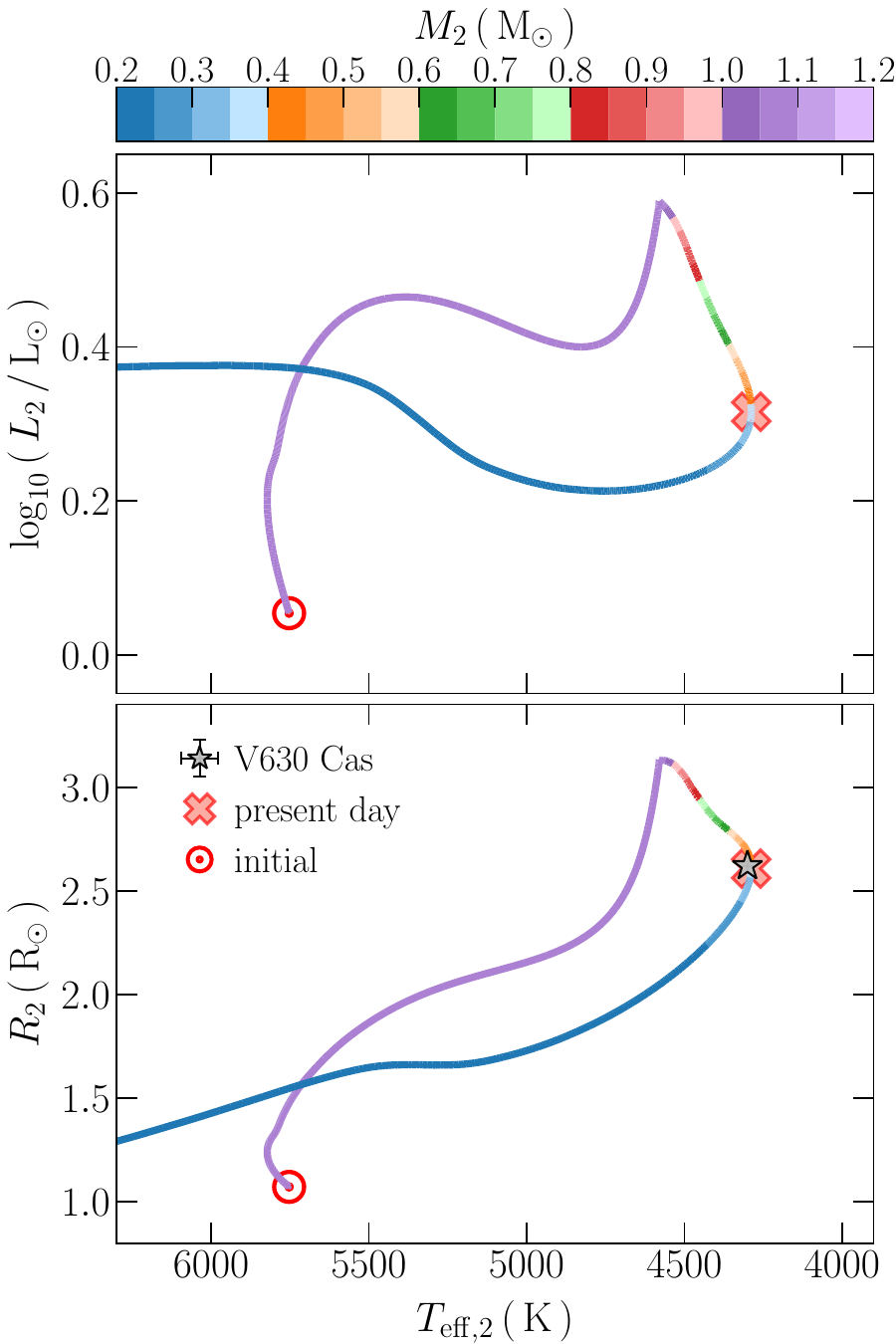}
\hspace{0.5cm}
\includegraphics[width=0.35\linewidth]{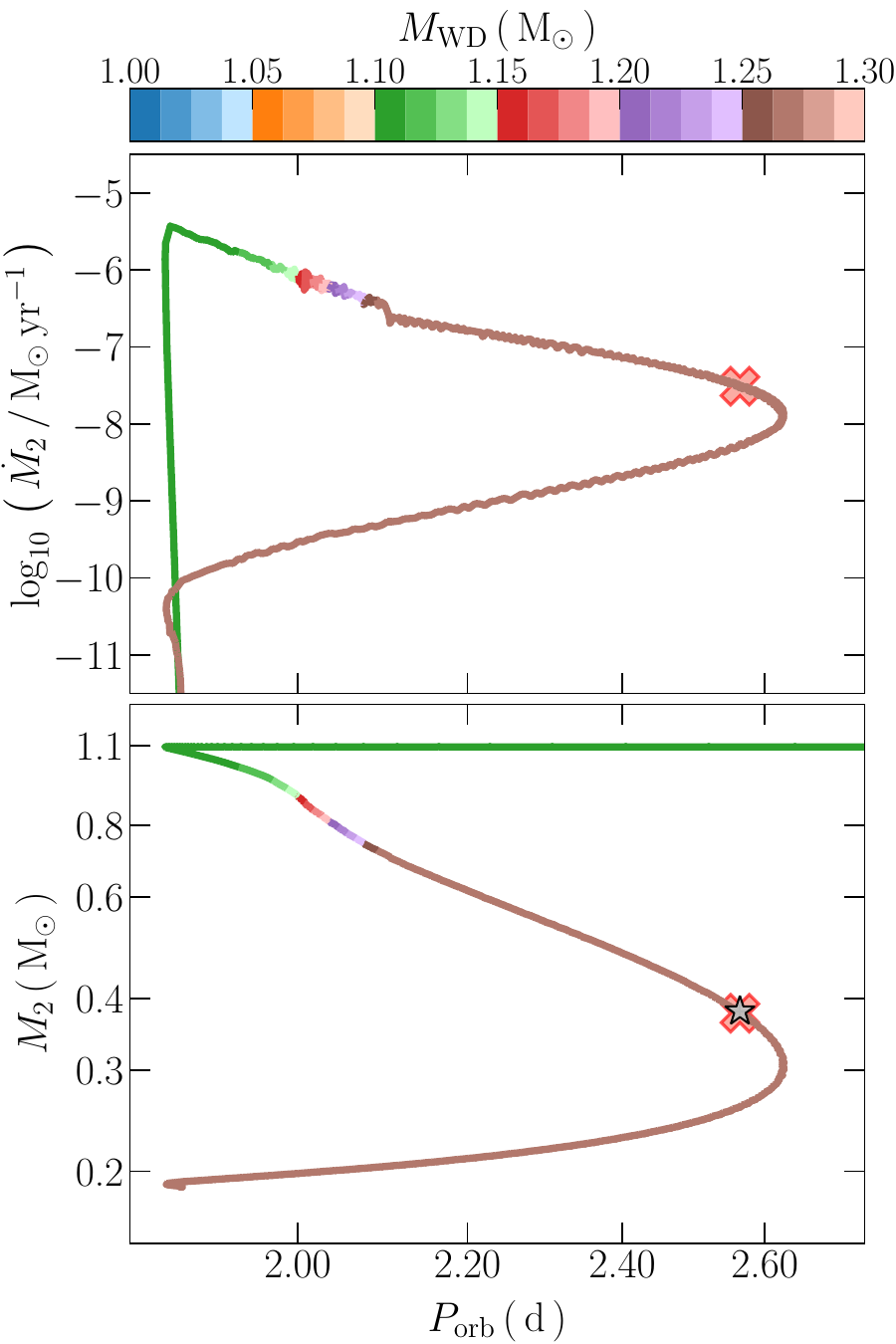}
\caption{Donor luminosity ($L_2$) and radius ($R_2$) against its effective temperature ($T_{\rm eff,2}$) (left panel) and evolution with the orbital period ($P_{\rm orb}$) of the mass transfer rate ($\dot{M}_2$) and the donor mass ($M_2$) (right panel) during post-CE evolution. The tracks are colour-coded by $M_2$ (left panel) and WD mass $M_{\rm WD}$ (right panel). The red solar symbol corresponds to the WD formation, and the thick red cross to the properties at the observed orbital period. During the detached phase, the orbital period decreases, and the WD companion evolves to a red giant star and expands before finally filling its Roche lobe. During the early CV phase, magnetic braking is sufficiently strong to drive the donor out of thermal equilibrium, and the WD mass increases due to the very high mass transfer rate. When the donor restores thermal equilibrium, the evolution becomes driven by the donor nuclear evolution and the WD no longer increases in mass. The orbital period increases during the entire evolution, except during late evolution, when magnetic braking becomes sufficiently strong to make the evolution converge.}
\label{fig:evol:RadLumTeffPorbMassMdot}
\end{figure*}

We allowed the stars to lose mass through winds, adopting the \citet{Reimers_1975} prescription, setting the wind efficiency to $0.5$. For the evolutionary phases with convective core, that is, core hydrogen and helium burning, we took into account exponential diffusive overshooting, assuming a smooth transition in the range ${1.2-2.0}$~\Msun~\citep[e.g.,][]{Anders_2023}. We assumed that the extent of the overshoot region corresponds to ${0.016~H_{\rm p}}$ \citep[e.g.,][]{Schaller_1992,Freytag_1996,Herwig_2000}, with $H_{\rm p}$ being the pressure scale height at the convective boundary. We treated convective regions using the scheme by \citet{Henyey_1965} for the mixing length theory, varying the mixing length (in units of $H_{\rm p}$): $1.5,2.0,2.5$ \citep[e.g.,][]{Joyce_2023}.

We assume as an initial configuration a detached post-CE binary consisting of a point-mass WD and a zero-age main-sequence star. The initial parameters of the post-CE binaries include WD masses of $0.8$, $0.9$, $1.0$, and $1.1$~\Msun, companion masses of $1.0$, $1.1$, $1.2$, and $1.3$~\Msun, and orbital periods ranging from $2$ to $15$~d, in steps of $0.1$~d. Initially, the initial mass of the WD might seem high for a post-CE binary. However, \citet{Adamane_2026} show that there are double DW (DWD) candidates within that mass range in the SDSS-V sample, so this initial range of masses is admissible. Furthermore, recent studies have investigated objects with very massive WDs, with masses greater than 1.2 \msun. For example, \citet{Yamaguchi_2024} presents observations of post-common envelope binaries (PCEBs) containing ultramassive WDs ($M_{WD} > 1.2$ \msun) and MS stars, with significantly wider orbits than previously known, with orbital periods of 18 to 49 days. Given that the typical timescale for the formation of an average CV WD \citep[${\sim300}$~Myr,][]{Belloni_2018b,Belloni_2020a} and the CE time we found in our pre-CE modeling with the BSE code (Table~\ref{Tab:FormationChannel}) are much shorter than the main-sequence lifetimes of our WD companions, starting the \textsc{MESA} simulations with a zero-age main-sequence star and ignoring pre-CE evolution does not affect our results. Despite that, after finding a reasonable pre-CE model with BSE, we reran the MESA model taking into account the companion age at the onset of WD formation. Regarding orbital angular momentum loss (AML), we assumed emission of gravitational waves, magnetic braking following the CARB prescription \citep{Van+Ivanova_2019}, and consequential AML due to mass loss from the system due to nova eruptions \citep{King_1995}. The Roche-lobe radius of the WD companion was computed using the fit of \citet{Eggleton_1983}. The mass transfer rates due to Roche-lobe overflow are determined following the atmospheric Roche-lobe overflow model put forward by \citet{Ritter_1988}.

We enforced that the WD accretor cannot always accrete all incoming mass. Depending on the accretion rate onto the WD, hydrogen shell burning could be stable. We implemented the critical accretion rate calculated by \citet[][]{Wolf_2013}, above which WDs are thermally stable, that is, hydrogen burns steadily in a shell. For accretion rates lower than this critical value, the WD undergoes nova eruptions, such that all of the accreted mass is assumed to be expelled from the binary. We further assumed that there is a maximum possible accretion rate \citep[][]{Wolf_2013} such that WDs accreting at rates above it will burn stably at this maximum rate, and the remaining non accreted matter will be piled up, forming a red giant-like envelope, which is assumed to be lost from the binary in the form of stellar-like winds. When hydrogen shell burning is stable, the helium burning shell underneath it could also be partially stable, leading to an increase in its mass. For helium accretion, i.e., the helium produced by stable hydrogen burning, we estimated the amount, which depends on the rate at which hydrogen is converted to helium \citep[e.g.][]{Tang_2024}, adopting the models from \citet{Kato_2004}.

Given the number of input parameters and assumptions summarised in Table~\ref{Tab:Assumptions}, it is worth explicitly stating how sensitive our conclusions are to them and to what extent the specific evolutionary track presented below is unique. Varying the mixing length, overshooting extent, metallicity, and opacity tables within the ranges explored in our grid changes the predicted donor temperature and radius at the observed orbital period. These choices are not the limiting factor. The dominant sensitivity is instead due to the angular-momentum-loss prescription: as discussed in Section~\ref{sec:evolution} below, only strong magnetic braking makes the evolution converge, while weaker magnetic braking fails to reproduce the properties of systems with evolved donor stars. Regarding uniqueness: our grid search over initial WD mass, companion mass, and orbital period (Table~\ref{Tab:Assumptions}) shows that several different combinations of these three parameters converge, at the observed orbital period, onto donor properties that agree with the SED-based $M_2$ and $R_2$ to within their uncertainties; we do not attempt a formal Bayesian inversion over the full parameter space here, and different initial conditions combined with different, similarly defensible, physical assumptions could in principle reach a comparable present-day configuration through a different evolutionary history. The model presented below should therefore be read as a demonstration of physical plausibility under the CARB prescription, not as a uniquely determined evolutionary history.

We provide in Table~\ref{Tab:FormationChannel} an example of an initial zero/age main-sequence binary that evolves into a binary with properties comparable to those of \vcas. The initial masses of the primary, which will become the WD, and the secondary, which will become the donor in \vcas, are $5.812$ and $1.045$~\Msun, respectively, while the initial orbital period is $6050$~d. The primary quickly evolves to an asymptotic giant branch star (within ${\sim85}$~Myr) and starts transferring mass to the secondary through wind accretion. As the primary further expands and the orbit shrinks due to strong tides, the primary eventually fills its Roche lobe at the beginning of the thermally pulsing phase. At this moment, the secondary mass has increased to $1.1$~\Msun, and the mass of the primary helium core (i.e. the future WD) has grown to $1.1$~\Msun.

After CE envelope evolution, the masses of the WD and its companion are $1.1$ and $1.1$~\Msun, respectively, and the orbital period is $12.4$~d. The orbital period is initially too long for the WD companion to fill its Roche lobe as a main-sequence star. As a matter of fact, it first became a subgiant and later a red giant, before filling its Roche lobe due to a combination of radius increase driven by stellar evolution and Roche lobe decrease driven by magnetic braking, as illustrated in the left panel of Figure~\ref{fig:evol:RadLumTeffPorbMassMdot}. Indeed, as soon as the WD companion becomes a subgiant, magnetic braking determines the angular momentum loss rates, which become very high when the WD companion becomes a red giant, reaching ${\sim10^{38}}$~dyn\,cm at the onset of Roche lobe overflow. This detached post-CE evolution takes around 8 Gyr (Table~\ref{Tab:FormationChannel}).

During early mass transfer evolution, as the donor is a red giant, the orbit expands in response to mass transfer. In addition, the timescale of mass loss driven by magnetic braking becomes shorter than the thermal timescale of the donor star, driving the donor star out of thermal equilibrium. Because of that, the mass transfer rate is initially very high (${\gtrsim10^{-6}}$~\Msunyr), which causes most of the accreted matter to pile up above the WD surface and be lost in the form of a fast wind. However, as soon as the mass transfer rate drops and reaches the range in which most hydrogen-rich accreted matter is burnt (${\sim10^{-7}-10^{-6}}$~\Msunyr), a large fraction of helium burning remains steady (${\sim70-80}$\,\%), causing an increase in the WD mass: it grows through accretion from $1.1$ to $1.274$~\Msun\ (Table~\ref{Tab:FormationChannel}), in excellent agreement with the self-consistently derived value of $M_1 = 1.27^{+0.10}_{-0.06}$\,\msun\ (Section~\ref{subsec:MassFuncWD}), as also illustrated in Figure~\ref{fig:evol:RadLumTeffPorbMassMdot}.

The thermal timescale mass transfer ends when the mass transfer rate drops below the critical rate for stable hydrogen burning at a donor mass of ${\sim0.65}$~\Msun. Therefore, at this point, the mass-loss timescale has sufficiently increased and has become greater than the thermal timescale of the donor star. From this point onward, the WD mass no longer increases due to cyclic hydrogen nova eruptions. Additionally, the evolution becomes entirely driven by the nuclear evolution of the donor star, as magnetic braking is not strong enough to make the mass-loss timescale shorter than the nuclear timescale of the donor star.

As the donor mass keeps decreasing, the mass transfer rate also decreases, while the orbit expands following the expansion of the donor star. \vcas is exactly in this phase of nuclear timescale evolution, and at the observed orbital period, the binary has properties comparable to those observed.

The future evolution of \vcas is similar until the trend in the orbital period evolution is reversed. This will happen when sufficient envelope mass is stripped from the donor star, making its nuclear timescale longer than the magnetic-braking-driven mass-loss timescale, at a donor mass of ${\sim0.3}$~\Msun. After this point, the orbital period decreases, and the donor star starts moving along the WD cooling sequence. When nearly all the envelope has been removed, the binary detaches, and the WD companion becomes a proto-WD.

\section{Discussion}
\label{sec:discuss}

\vcas was discovered more than half a century ago, and despite exhibiting typical CV signatures in both its spectrum and light curve, it remained an exceptional case due to its unusually long orbital period, rivaled only by GK\,Per for many decades. However, in contrast to the substantial attention devoted to other long-period CVs, \vcas has been the subject of only one detailed observational study by \citet{Orosz_2001}. Motivated by the evolutionary implications, we conducted a new observational campaign to revisit this system, combining time-resolved spectroscopy with high-cadence photometry from \textsc{TESS}, and taking advantage of the precise distance measurement now available from \textsc{Gaia}.

Our observations largely confirm the phenomenology described by \citet{Orosz_2001}, but reveal several new features. First, the \textsc{TESS} light curves obtained during Sectors 17 and 84 show variability morphologies that differ from one to the other. In Sector 84, the modulation resembles that previously published, with the deeper minimum at phase 0.5 as expected for ellipsoidal variability with disc shading of the donor. The Sector 17 light curve, however, exhibits an asymmetric double-humped morphology with a displaced dip, indicating an epoch-dependent non-axisymmetric structure in the accretion disc. The \textsc{AAVSO} long-term record shows a significant dimming episode prior to the Sector 17 epoch, suggesting a changed disc state. We attribute the variable light-curve morphology to changes in the disc brightness distribution—possibly reflecting a precessing or eccentric disc—rather than a change in the orbital geometry.

More critically, our spectroscopic data reveal that the H$\alpha$ emission line exhibits a complex, phase-variable profile that can formally be decomposed into two Gaussian components. This complicates the interpretation of the emission-line radial velocity curves, which are often used as a proxy for the WD's motion. Although \citet{Orosz_2001} avoided using these velocities for dynamical modeling, they nonetheless derived a semi-amplitude $K_1$ based on a single-Gaussian fit.

The Doppler tomography of H$\alpha$ (Figure~\ref{fig:DopplerMap}) does not show the compact spot structure expected from a classical hot-spot or S-wave: the emission is distributed in a broad, azimuthally asymmetric pattern centered on the WD, with maximum intensity at blueshifted velocities near $\phi \approx 0.25$ rather than at the predicted disc--stream impact location. We therefore interpret the H$\alpha$ emission as arising primarily from the optically thin outer rim of a large accretion disc, which is azimuthally asymmetric—possibly due to an eccentric or precessing disc structure, as suggested by the epoch-dependent variability seen in the \textsc{TESS} light curves. The double-peaked profile morphology and its orbital variation are natural consequences of such a geometry rather than evidence for two kinematically distinct point sources. The double Gaussian decomposition (components A and B) provides a useful parameterization of the changing line asymmetry, but the components should not be over-interpreted as two physically separate emission sites. The complex systemic velocities and large amplitudes of both components, compared to the independently determined $K_2 = 132.9$\,\kms\ from absorption lines, are consistent with disc emission modulated by a rotating asymmetric emissivity pattern. Our findings highlight the complexity of emission line formation in long-period CVs with large, evolved accretion discs, and underscore that Balmer emission velocities alone are insufficient for dynamical mass determinations in these systems.

Most importantly, we obtained a constraint on the mass of the donor star, which is crucial for the evolutionary considerations. Ambiguities remain—in the donor contribution to the total flux, the bolometric correction, and the distance—but we can confidently state that the donor mass is higher than estimated by \citet{Orosz_2001} and lower than that of a regular ZAMS K4--K5 star. The best photometric estimate is $M_2 = 0.38$\,\msun\ at the adopted distance $D = 2409$\,pc, with the quoted range $M_2 = 0.30$--$0.53$\,\msun\ driven primarily by the Bailer-Jones distance uncertainty. Additional systematic contributions arise from the assumed donor flux fraction (${\approx}50$\,per cent of the optical flux), the adopted $T_{\rm eff} = 4300$\,K (SED solutions span 4100--4400\,K), and the bolometric correction ($BC = -0.65$, uncertain by ${\sim}0.1$\,mag). These sources are mutually independent and unlikely to bias the result coherently in the same direction; partial cancellation is expected, so the net additional uncertainty is smaller than a naive linear sum of the individual contributions.

\section{CONCLUSIONS}
\label{sec:conclusions}

We have presented a comprehensive observational and theoretical study of the long-period CV \vcas\ ($P_{\rm orb} \approx 2.564$\,d), combining time-resolved optical spectroscopy obtained at the OAN-SPM over 22 observing runs, four sectors of \textsc{TESS} photometry, and binary evolution models computed with \textsc{MESA} under the CARB magnetic braking prescription \citep{Van+Ivanova_2019}. Our main conclusions are the following.

\begin{enumerate}

\item Refined orbital ephemeris.
Using the zero-point epoch of \citet{Thorstensen_2017} and a baseline of ${\sim}3800$ orbital cycles spanning our spectroscopy and archival \textsc{TESS} data, we refine the orbital period to $P_{\rm orb} = 2.563886 \pm 0.000003$\,d, extending the precision by one significant digit relative to the \citet{Thorstensen_2017} determination of $2.56388(2)$\,d.

\item Binary system parameters from \textsc{Gaia} astrometry.
Adopting the donor semi-amplitude $K_2 = 132.9 \pm 4.0$\,\kms\ from \citet{Orosz_2001}, the \textsc{Gaia} geometric distance $D = 2409$\,pc \citep{Bailer-Jones_2021}, and the self-consistently derived inclination $i \simeq 70^{\circ}$ (Section~\ref{subsec:MassFuncWD}), required for consistency with the mass function $f(M_1) = 0.62$\,\msun\ and the absence of an eclipse, we derive a WD mass $M_1 \approx \textbf{1.27}$\,\msun. The donor's luminosity and colour at the \textsc{Gaia} distance place it on the donor diagnostic diagram \citep{Tovmassian_2025} with a best-estimate mass $M_2 = 0.38$\,\msun\ and radius $R_2 = 2.62$\,\rsun; the Bailer-Jones distance uncertainty maps to the range $M_2 = 0.30$--$0.53$\,\msun\ ($R_2 = 2.41$--$2.95$\,\rsun). These values confirm that the donor is a significantly undermassive, thermally inflated subgiant-to-red-giant, with a mass substantially higher than the $M_2 < 0.2$\,\msun\ estimated by \citet{Orosz_2001}, who lacked a reliable distance.

\item Emission-line velocities trace disc structure, not orbital motion.
The H$\alpha$ emission profile of \vcas\ is predominantly single-peaked despite $i \approx \textbf{70}^{\circ}$, exhibiting complex, phase-variable asymmetry that can be decomposed into two Gaussian components. However, both components yield nearly equal semi-amplitudes ($K_{\rm A} \approx K_{\rm B} \approx 118$\,\kms) in strict anti-phase ($\Delta\phi \approx 0.46$), with apparent systemic velocities that differ substantially from each other and from the true $\gamma = -93.5$\,\kms\ independently established from absorption lines. Doppler tomography confirms that H$\alpha$ emission arises from a large, azimuthally asymmetric accretion disc with no compact hot-spot; the velocity variations are a consequence of a rotating non-axisymmetric emissivity pattern in the disc rim, not of two kinematically distinct emitters. We conclude that Balmer emission line velocities in \vcas\ are not a reliable tracer of the WDs orbital motion. This behaviour — disc structure masquerading as orbital velocity — appears to be a characteristic feature of long-period CVs harbouring large, geometrically extended discs, as observed in V479\,And, V1082\,Sgr \citep{Tovmassian_2025}, and SDSS\,J085210.48+783246.6 \citep{Tovmassian_2026}.

\item Evolutionary status: nuclear-timescale mass transfer.
\textsc{MESA} models with CARB magnetic braking reproduce the observed properties of \vcas\ through a progenitor consisting of a $1.1$\,\msun\ WD and a $1.1$\,\msun\ main-sequence companion in a $12.4$\,d post-CE orbit. After ${\sim}8$\,Gyr of detached evolution, the companion filled its Roche lobe as a red giant. A brief (${\sim}3$\,Myr) episode of thermal-timescale mass transfer then grew the WD from $1.1$ to $1.274$\,\msun, consistent with our spectroscopic constraint. \vcas\ is currently in the nuclear-timescale phase, in which the orbital period increases in response to mass transfer driven by the slow expansion of the donor's helium core. The future evolution will continue in this direction until magnetic braking becomes strong enough to reverse the period increase at $M_2 \approx 0.30$\,\msun, after which the system will evolve toward shorter periods and the donor will eventually detach as a proto-WD.

\item The pivotal role of the donor mass and radius.
The evolutionary model that matches the orbital period and WD mass of \vcas\ predicts $M_2 = 0.38$\,\msun, $R_2 = 2.61$\,\rsun, and $T_{\rm eff,2} = 4288$\,K at the observed period — in good agreement with our photometric constraint. This match is non-trivial: models that use standard, weaker magnetic braking prescriptions fail to reproduce either the onset of Roche-lobe overflow at $P_{\rm orb} \approx 2.5$\,d or the current donor properties. Pinning down $M_2$ and $R_2$ is therefore essential not only to characterise the system, but to discriminate between competing angular-momentum-loss prescriptions. This was only possible here because of the precise \textsc{Gaia} parallax. We argue that accurate distance-dependent donor mass and radius determinations are the critical observational input for testing magnetic braking models in long-period CVs, a conclusion that generalises to the growing sample of similar systems \citep{Tovmassian_2025,Tovmassian_2026}.

\end{enumerate}


\bibliography{references} 

\section{ACKNOWLEDGEMENTS}
We thank the anonymous reviewer for the careful reading and constructive suggestions that led to a significant improvement of the manuscript. GT was supported by grants  IN108426 from the Programa de Apoyo a Proyectos de Investigaci\'on e Innovaci\'on Tecnol\'ogica (DGAPA-PAPIIT). IMZ acknowledges support from CONAHCYT grant 1047189. This research was supported in part by NSF grant PHY-1748958 to the Kavli Institute for Theoretical Physics (KITP), which provided us with an opportunity to forge a collaboration that led to this research. This work is supported by the National Natural Science Foundation of China (NSFC, Nos. 12288102, 12125303), the Strategic Priority Research Program of the Chinese Academy of Sciences (grant Nos. XDB1160303, XDB1160300, XDB1160000, XDB1160200, XDB1160201), the CAS Project for Young Scientists in Basic Research (YSBR-148), the Yunnan Revitalization Talent Support Program ``YunLing Scholar'' project, International Centre of Supernovae (ICESUN), Yunnan Key Laboratory of Supernova Research (No. 202505AV340004),  the New Cornerstone Science Foundation through the XPLORER PRIZE, and the Yunnan Revitalization Talent Support Program-Science \& Technology Champion Project (No. 202305AB350003). This paper includes data collected with the \textsc{TESS} mission, obtained from the MAST data archive at the Space Telescope Science Institute (STScI). Funding for the \textsc{TESS} mission is provided by the NASA Explorer Program. STScI is operated by the Association of Universities for Research in Astronomy, Inc., under NASA contract NAS 5-26555. Support for MAST for non-HST data is provided by the NASA Office of Space Science via grant NAG5-7584 and by other grants and contracts. 

\end{document}